\documentclass[preprint,12pt]{aastex}
\usepackage{natbib}

\newcommand{\h}{\mbox{$^h$}}
\newcommand{\m}{\mbox{$^m$}}
\newcommand{\s}{\mbox{$^s$}}
\newcommand{\degree}{\mbox{$^{\circ}$}}
\newcommand{\am}{\mbox{\arcmin}}

\newcommand{\kms}{\mbox{km s$^{-1}$}}% km/s

\newcommand{\lsun}{\mbox{L$_\odot$}}% Lsun
\newcommand{\msun}{\mbox{M$_\odot$}}% Msun

\input{epsf}

\begin{document}

%%%%%%%%%%%%%%%%%% title %%%%%%%%%%%%%%%%%%%%%%%%%%%%%%%%%%%%%%%%
\title {\bf The Spitzer c2d Survey of Large, Nearby, Interstellar Clouds. I. Chamaeleon II Observed with MIPS}
\author{Kaisa E. Young\altaffilmark{1},
Paul M. Harvey\altaffilmark{1},
Timothy Y. Brooke\altaffilmark{2},
Nicholas Chapman\altaffilmark{3},
Jens Kauffmann\altaffilmark{4}, 
Frank Bertoldi\altaffilmark{5},
Shih-Ping Lai\altaffilmark{3},
Juan Alcal\'{a}\altaffilmark{6},
Tyler L. Bourke\altaffilmark{7},
William Spiesman\altaffilmark{1},
Lori E. Allen\altaffilmark{7},
Geoffrey A. Blake\altaffilmark{8},
Neal J. Evans II\altaffilmark{1},
David W. Koerner\altaffilmark{9},
Lee G. Mundy\altaffilmark{3},
Philip C. Myers\altaffilmark{7},
Deborah L. Padgett\altaffilmark{10},
Anandi Salinas\altaffilmark{1},
Anneila I. Sargent\altaffilmark{2},
Karl R. Stapelfeldt\altaffilmark{11},
Peter Teuben\altaffilmark{3},
Ewine F. van Dishoeck\altaffilmark{12},
Zahed Wahhaj\altaffilmark{9}}

\altaffiltext{1}{Astronomy Department, University of Texas at Austin, 1 University Station C1400, Austin, TX 78712-0259; kaisa@astro.as.utexas.edu, pmh@astro.as.utexas.edu, spies@astro.as.utexas.edu, nje@astro.as.utexas.edu}
\altaffiltext{2}{Division of Physics, Mathematics, \& Astronomy, MS 105-24, California Institute of Technology, Pasadena, CA 91125; tyb@astro.caltech.edu, afs@astro.caltech.edu}
\altaffiltext{3}{Astronomy Department, University of Maryland, College Park, MD 20742; chapman@astro.umd.edu, slai@astro.umd.edu, lgm@astro.umd.edu}
\altaffiltext{4}{Max-Planck-Institut f\"{u}r Radioastronomie, Auf den H\"{u}gel 69, D-53121 Bonn, Germany; jkauffma@mpifr-bonn.mpg.de}
\altaffiltext{5}{Radioastronomisches Institut der Universit\"{a}t Bonn, Auf den H\"{u}gel 71, D-53121 Bonn, Germany; bertoldi@uni-bonn.de}
\altaffiltext{6}{INAF - Osservatorio Astronomico di Capodimonte, via Moiariello 16, 80131 Napoli, Italy; jmae@sun1.na.astro.it}
\altaffiltext{7}{Smithsonian Astrophysical Observatory, 60 Garden Street, MS42, Cambridge, MA 02138; leallen@cfa.harvard.edu, tbourke@cfa.harvard.edu, pmyers@cfa.harvard.edu}
\altaffiltext{8}{Division of Geological and Planetary Sciences, MS 150-21, California Institute of Technology, Pasadena, CA 91125; gab@gps.caltech.edu}
\altaffiltext{9}{Northern Arizona University, Department of Physics and Astronomy, Box 6010, Flagstaff, AZ 86011-6010; koerner@physics.nau.edu, zwahhaj@physics.nau.edu}
\altaffiltext{10}{Spitzer Science Center, MC 220-6, Pasadena, CA 91125; dlp@ipac.caltech.edu}
\altaffiltext{11}{Jet Propulsion Laboratory, MS 183-900, California Institute of Technology, 4800 Oak Grove Drive, Pasadena, CA 91109; krs@exoplanet.jpl.nasa.gov}
\altaffiltext{12}{Leiden Observatory, Postbus 9513, 2300 RA Leiden, Netherlands; ewine@strw.leidenuniv.nl}

%%%%%%%%%%%%%%%%%%%% Abstract %%%%%%%%%%%%%%%%%%%%%%%%%%%%%
\begin{abstract}

We present maps of over 1.5 square degrees in Chamaeleon (Cha) II
at 24, 70, and 160 \micron\ observed with the Spitzer Space Telescope
Multiband Imaging Photometer for Spitzer (MIPS) and a 1.2 square
degree millimeter map from SIMBA on the Swedish-ESO Submillimetre
Telescope (SEST). The c2d Spitzer Legacy Team's data reduction
pipeline is described in detail. Over 1500 24 \micron\ sources and 41
70 \micron\ sources were detected by MIPS with fluxes greater than
10-$\sigma$. More than 40 potential YSOs are identified with a MIPS and
2MASS color-color diagram and by their spectral indices, including two
previously unknown sources with 24 \micron\ excesses. Our new SIMBA
millimeter map of Cha II shows that only a small fraction of the gas
is in compact structures with high column densities. The extended
emission seen by MIPS is compared with previous CO observations. Some
selected interesting sources, including two detected at 1 mm,
associated with Cha II are discussed in detail and their SEDs
presented. The classification of these sources using MIPS data is
found to be consistent with previous studies.

\end{abstract}


%%%%%%%%%%%%%%%%%%% Main text %%%%%%%%%%%%%%%%%%%%%%%%%%%%

\section{Introduction}

The Chamaeleon (Cha) II molecular cloud is one of five clouds that
were mapped with the Spitzer Space Telescope (SST) as part of the
``From Molecular Cores to Planet-forming Disks'' Legacy project
\citep{eva03}. The main goal of the ``Cores to Disks'' (c2d) Legacy
project is to study the evolution of star and planet formation from
cold molecular cores to protoplanetary disks in a range of
environments. The project makes use of all of the instruments on the
SST (the Multiband Imaging Photometer for Spitzer (MIPS), the InfraRed
Array Camera (IRAC), and the InfraRed Spectrograph (IRS)) to observe
nearby molecular clouds, isolated regions of star formation, and young
stellar objects (YSOs), defined broadly in this work as sources with
infrared excesses. One goal of the c2d project is to survey the YSO
population of the five large molecular clouds for new candidate
protostars as well as substellar objects down to 0.001 \lsun.  This
work describes MIPS observations of one of the molecular clouds mapped
by c2d, Cha II, and is the first of a series of papers that present
initial results from the c2d observations. We also present a new
millimeter map of Cha II from SIMBA on the Swedish-ESO Submillimetre
Telescope (SEST). The IRAC observations of Cha II will be described by
Porras et al.\ (2005, in preparation). Later papers will follow that
combine the data from all of the SST observations as well as
complementary observations at other wavelengths and include further
analysis of the YSOs and cloud environments.

Chamaeleon is a star forming region in the Southern sky that includes
at least half a dozen distinct clouds \citep{bou98}. Cha II ($\alpha$
= 13\h, $\delta$ = $-$77\degree; J2000), at a distance of 178 $\pm$ 18
pc \citep{whi97}, is known to harbor dense molecular cores
\citep{miz99}, T Tauri stars, and YSOs \citep{sch77, whi91, hug92,
pru92, vuo01}. Pointed ROSAT observations revealed many X-ray sources
in Cha II including weak-line and classical T Tauri stars
\citep{alc00}. In addition to low-mass young stars, one intermediate
mass Ae star, DK Cha (IRAS 12496-7650), and the well-studied
Herbig-Haro object HH 54 (e.g., \citealt{kne92, lis96, neu98}) are
also associated with Cha II. \citet{per03} studied a portion of Cha II
(28\arcmin\ $\times$ 26\arcmin) near the Class I source IRAS
12553-7651 (ISO-ChaII-28) in the mid-infrared (6.75 and 14.3 \micron)
with ISOCAM on the ISO satellite as well as with near-infrared images
and found 4 new candiate YSOs (see their Table 4) in the observed
region. Our MIPS survey of Cha II covers more area (about 7 times) at
longer wavelengths (24, 70, and 160 \micron) than the ISO observations
with greater sensitivity (1-$\sigma$ $\approx$ 0.15 mJy at 24 \micron\
vs.\ 1.3 mJy at 14 \micron).

The five clouds in the c2d survey were chosen to span a range of star
formation conditions.  Cha II is one of the most quiescent of the
clouds, with only a few known very young (Class I) objects. Most of
the more than 30 previously identified YSOs in Cha II are T Tauri
stars. Cha II is a typical site of low-mass star formation compared
with other regions, with a star formation efficiency ($\sim$1\%;
\citealt{miz99}) similar to that of Taurus (2\%; \citealt{miz95}) and
the Ophiuchus-north region (0.3\%; {\it l} = 355 -- 12\degree, {\it b}
= 15 -- 23\degree; \citealt{tac00}). \citet{bou98} studied Cha II with
CO and IRAS and reported that the ratio between the mass needed to
gravitationally bind the cloud and the gas mass is 7.8, indicating
that Cha II is not gravitationally bound. They suggest that Cha II may
be two independent clouds not bound together, because the CO spectrum
has two lines. \citet{miz99} mapped Cha II in C$^{18}$O and combined
their work with previous infrared and optical studies of the
cloud. They report two velocity components in the C$^{18}$O for the
southeastern part of the cloud and suggest that Cha II and Cha III
might be superimposed along the line of sight. Therefore, it is
unclear if Cha II itself is bound. \citet{miz99} also compared their
Cha II results with similar previous work on other molecular clouds,
such as Taurus \citep{oni96} and the Ophiuchus-north region
\citep{tac00} (see \citet{miz99} Table 2). The mean molecular column
density ($N$(H$_2$)) of Cha II, $5.1 \times 10^{21}$ cm$^{-2}$, is
slightly smaller than that of Ophiuchus-north, $5.9 \times 10^{21}$
cm$^{-2}$, or Taurus, $6.9 \times 10^{21}$ cm$^{-2}$ (\citealt{miz99}
Table 2).  \citet{miz99} also found a half-power Gaussian linewidth,
$\Delta V$, of 0.78 \kms\ in Cha II, which is similar to
Ophiuchus-north ($\Delta V$ = 0.7 \kms) but greater than Taurus
($\Delta V$ = 0.49 \kms).  \citet{miz99} report that the young stars
in Cha II are not densely clustered, but in sparse groups. There is a
group of T Tauri stars in the southern tip of the cloud, and other
known T Tauri stars are scattered along the eastern edge and to the
east of the cloud \citep{miz99}.

\section{Observations}\label{obs}

\subsection{MIPS Observations}

The general Spitzer mapping strategy of Cha II was consistent with
that of the other four molecular clouds in the c2d Legacy program
\citep{eva03}. All of the area in the Cha II molecular cloud with an
A$_V$ greater than 2 magnitudes \citep{cam99} was mapped with the
Multiband Imaging Photometer for Spitzer (MIPS; \citealt{rie04}) on 06
April 2004 (PID: 176, AOR keys: 0005741056, 0005744640, 0005744896,
0005745152).  A total area of over 1.5 square degrees was mapped in
MIPS fast-scan mode. The fast-scan mode observes in all MIPS
wavelength bands (24, 70, and 160 \micron) simultaneously. The pixel
size is 2.5\arcsec\ at 24 \micron, 10\arcsec\ at 70 \micron, and
16\arcsec\ $\times$ 18\arcsec\ at 160 \micron. A more detailed
desciption of the MIPS instrument can be found in \citet{rie04} or in
Chapter 8 of the \citet{spi04}. Cha II was covered by two Spitzer
astronomical observation requests (AORs), 1\degree\ long in the scan
direction by 11 scan legs (0.7\degree) and 14 scan legs (0.9\degree)
wide. The scan legs were offset from each other by 240\arcsec.  Figure
\ref{av_yso} shows the 24 \micron\ map of Cha II; the scan direction
is approximately East-West.  In fast-scan mode, MIPS integrates for 3
seconds per pointing on the sky. Any individual position on the sky is
observed 5 times at 24 and 70 \micron\ for an integration time of 15
seconds per position on the sky. However, the effective 70 \micron\
field of view (FOV) is only half that of the 24 \micron\ array (about
5\arcmin\ $\times$ 5\arcmin), because only one half of the 70 \micron\
array returns useable data. The 160 \micron\ coverage in fast-scan
mode is not complete by design. The 160 \micron\ array has a FOV of
only 0.5\arcmin\ $\times$ 5\arcmin\ and is further reduced by a dead
read-out \citep{rie04}. The integration time at any position on the
sky at 160 \micron\ is either 0 or 3 seconds in fast-scan mode. The
integration times are higher in all bands in regions where the scan
legs or AORs overlap.

The entire cloud was observed twice. The second observation, or epoch,
began about 6 hours after the first. The AORs were designed with a 3
to 6 hour separation of epochs to enable the detection and removal of
asteroids at 24 \micron. The two epochs were offset from one another
on the sky by about half an array (125\arcsec) in the cross scan
direction. The offset was designed to fill in the gaps left by the
side of the 70 \micron\ array that does not return usable data.  This
observation strategy results in approximately the same coverage at 24
and 70 \micron\ but less total (2-epoch) integration time at 70
\micron\ (15 seconds) than at 24 \micron\ (30 seconds). Additionally,
the second epoch was shifted by 80\arcsec\ in the scan direction to
partially fill the gaps in the 160 \micron\ coverage area. The 160
\micron\ coverage is still not complete with two observations, but the
gaps are minimized.

\subsection{SIMBA Observations}

Millimeter continuum observations were made 10 -- 15 November 2001
with the 37-channel bolometer array SIMBA (SEST Imaging Bolometer
Array) at the SEST on La Silla, Chile. We observed two regions which
cover the areas of highest optical extinction towards the Cha II
complex. The northern region has an approximate extent of $4000 \times
4000 ~ \rm arcsec^2$, and the southern one has an extent of roughly
$2000 \times 2000 ~ \rm arcsec^2$ for a total coverage of over 1.2
square degrees. The area mapped by SIMBA is slightly smaller than and
fits within the MIPS observation area. SIMBA is sensitive between
about $200$ and $280 ~ \rm GHz$, with an effective frequency of $250 ~
\rm GHz$ or an effective wavelength of $1.2 ~ \rm mm$.

In total, 46 maps were taken in the fast scanning mapping mode with a
scanning speed of $160 ~ \rm arcsec \, s^{-1}$.  Each of the maps
consists of 89 subscans in azimuth with a length of 1400\arcsec, with
16\arcsec\ of spacing in elevation between subscans (and thus a total
map extent of 1408\arcsec\ in elevation). The pointing uncertainties
are of order 7\arcsec\ to 16\arcsec, as determined from (and corrected
on the basis of) offsets between the SIMBA and MIPS positions of IRAS
12496-7650, while the SEST has a HPBW of 24\arcsec\ at $250 ~ \rm
GHz$. The zenith opacity, which was determined by skydips performed at
least every second hour for most of the data (and at least every fifth
hour in some cases), ranged from $0.23$ to $0.37$. Maps of Uranus and
Mars were used for calibration purposes. The derived calibration
factor of $123 ~ \rm mJy ~ counts^{-1}$ is in very good agreement with
the factor of $130 ~ \rm mJy ~ counts^{-1}$ determined by the
telescope crew for this observing period. Based on frequent
observations of $\eta$ Carin\ae{}, we estimate the relative
calibration uncertainty to be within $\pm$15\%.

\section{Data Reduction}\label{reduce}

\subsection{MIPS Data}

MIPS images were processed by the Spitzer Science Center (SSC) using
the standard pipeline (version S9.5) to produce Basic Calibrated Data
(BCD) images (\citealt{gor05a}; \citealt{mas04}; \citealt{mips04}).
The BCD images were then further corrected for some instrumental
signatures as described below.  Other instrumental signatures, e.g.,
dark latents, column and row pulldown effects, and streaks extending
from bright objects, have not been corrected. Bright latent images
occur at roughly 60\arcsec\ intervals in the scan direction after a
bright object. No attempt was made to explicitly remove cosmic rays or
bright latent images from the BCD images.  We used either the
redundancy and outlier rejection in mosaicking, or inspection, to
avoid mis-identifying these as point sources.  Some of the bright
latent images can be seen to the East and West of the bright sources
in the 24 \micron\ map of Cha II (Fig. \ref{av_yso}). Complete details
of the c2d processing are available from the SSC in the documentation
at {\it http://ssc.spitzer.caltech.edu/legacy/} \citep{eva04}.

\subsubsection{24 Micron Fast-scan Map Data}\label{24}

With the 24 $\mu$m array, a ``jailbar'' pattern results from bright
sources or from cosmic ray hits.  The jailbars are time-dependent
variations in the relative gains between readouts in the array. We
applied a multiplicative correction to each BCD frame for any
detectable fixed-amplitude jailbar pattern across the array, bringing
lower columns up to the level of the highest columns.  For some data,
we corrected the jailbarring in two sections, rows above and rows
below a bright object.  A few frames had a more complex jailbar
pattern and these were dropped from further analysis. Also, ``first
frame'' corrections were applied.  Scale factors were applied to the
first four frames in a map to bring them up to the median of
subsequent frames.  Finally, frames in each scan leg were
median-combined to create ``self-flats''.  Each frame was divided by
the self-flat for that scan leg.  This procedure corrected for
residual jailbarring and a 1 to 2\% gradient along the column
direction. Uncertainties resulting from these corrections to the pixel
values are estimated to be typically $\lesssim$ 0.2 \%, and the
resulting uncertainties in point-source fluxes $\lesssim$ 0.03 mJy.
This uncertainty is small compared to other sources of uncertainties.

\subsubsection{70 and 160 Micron Fast-scan Map Data}

The 70 and 160 \micron\ data contain several instrumental signatures.
The SSC provided two types of BCD images: normally processed
(unfiltered), and those with a time-median filter applied that removes
most of the background signal (filtered).  Following the current
recommendation of the SSC, we used the 70 \micron\ filtered data to
obtain point-source fluxes.  However, only the unfiltered data
preserve extended emission.
                                                                               
We are interested ultimately in fluxes for both point sources and
extended emission.  In this paper, we present preliminary maps made
from unfiltered 70 and 160 \micron\ data.  The 70 \micron\ map
includes approximate corrections for instrumental effects as described
below. The flux calibration for extended emission in the 70 and 160
\micron\ data is uncertain at the present time.  The corrections
applied at 70 \micron\ may not lead to properly calibrated data.  The
maps should be seen as only illustrating rough levels of extended
emission.
                
First, after stimulator (stim) flashes in 70 \micron\ data, there was
a ``stim latent'' signal present in subsequent BCD frames.  The first
frame after the stim flash was most affected; 4 frames after the stim
flash could have some residual effect.  We applied a column-by-column
subtractive correction based on the level after the latents decayed.
Second, the unfiltered data had striped patterns approximately along
the column direction due to suspected response variations (of order
several $\times$ 10\%).  We created an ``illumination correction''
made from the data themselves by median-combining frames in each scan
leg.  However, the corrected data are of uneven quality.  Surface
brightness uncertainties introduced by these corrections are difficult
to estimate without better understanding of the instrument.  However
the amplitude of the illumination corrections, several $\times$ 10\%,
may be a rough estimate.

No changes were made to 160 \micron\ unfiltered BCD data as the steps
needed for extended-source calibration are still in a preliminary
stage.

\subsubsection{Image Mosaicking}\label{mosaic}

\begin{figure}
\plotone{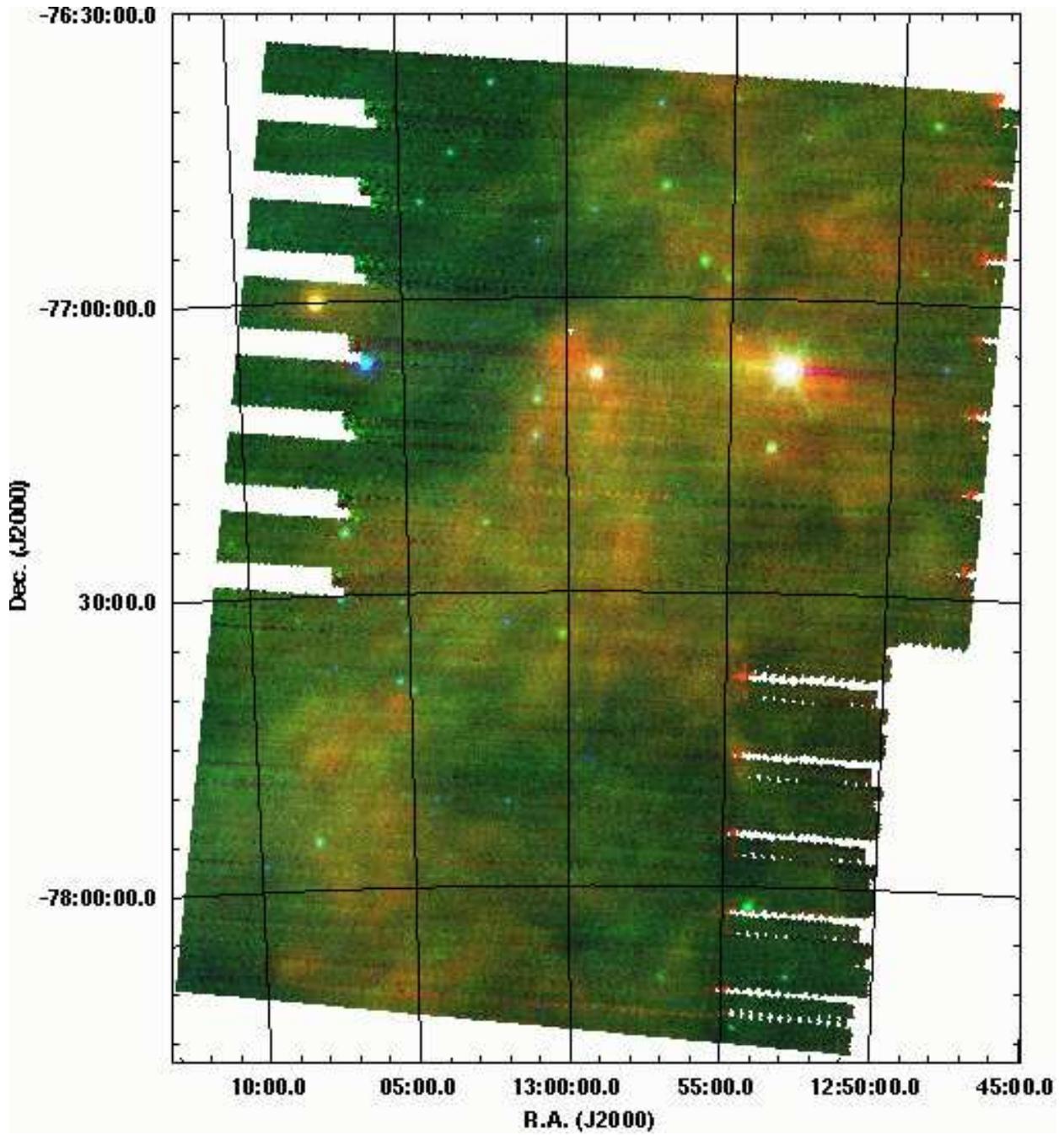} \figcaption{\label{map} 3-Color MIPS 24, 70, and
160 \micron\ (blue, green, and red, respectively) map of the Cha II molecular
cloud. The figure has been cropped to include only areas where there
are data at all 3 wavelengths.}
\end{figure}

The mosaic images produced for each wavelength are important both for
the overview they provide of the mapped area, especially of the
diffuse extended emission, and because our source extraction process
begins by searching for sources in the mosaics.  A three-color (24,
70, \& 160 \micron) mosaicked image of Cha II is shown in Figure
\ref{map}. This map shows the unfiltered 70 and 160 \micron\ data to
display the extended emission in Cha II. The primary tool for
mosaicking is the Spitzer Science Center's ``MOPEX'' code
\citep{mac04}.  This code includes a continuously evolving set of
modules that correct for some problems in the BCD and attempt to piece
many BCD images together into large mosaics.  Optional modules exist
to filter out radiation hits (outlier rejection) and to correct for
small positional errors in the FITS headers (position refinement).

For the 24 \micron\ data, we have used MOPEX with the outlier
rejection and position refinement modules turned on with nominal
parameters. In order to detect asteroids, we have divided the total
data set into two separate groups of BCD images, according to the two
epochs in which our data were acquired, and processed each
independently. The actual process of identifying asteroids and
combining the two epochs into a single, higher reliability source list
is described later in section \ref{bandmerge}. In total, we construct
mosaics for each epoch separately plus a combined mosaic in which the
highest signal-to-noise ratio is obtained.

As part of its standard processing, MOPEX makes use of mask files to
avoid including BCD pixels in the final mosaic that have been
identified as having some problem.  For our mosaics, we have used a
combination of the masks created in the initial c2d processing
together with the outlier masks constructed during the
outlier-rejection processing by MOPEX.  The masks created by the c2d
team effectively combine the bad pixel masks (``pmask'') and the bad
detection masks (``dmask'') of the SSC.

The geometry of the mosaics constructed with MOPEX for this paper was
such that we kept a 1:1 pixel size ratio between the basic instrument
pixel size and mosaic pixel size, and we assembled the BCD images in a
coordinate system essentially fixed with respect to the average
instrument rotation projected on the sky.  The former minimizes file
sizes and processing time.  The latter simplifies the process of point
source extraction from the mosaics since the instrumental PSF will
have the same rotational orientation for any mosaic.

\subsubsection{Source Extraction}\label{extract}

The process of source extraction involves finding likely compact
objects in the data and characterizing them by flux, position, and
some estimate of the way in which they might differ from a ``perfect''
point-like object.  The details of our point source extractor will be
described in a future paper (Harvey et al.\ 2005, in preparation), but
we summarize the main points here.

The source extraction tool, c2dphot, is based on the venerable
``DoPHOT'' code described by \citet{sch93}.  Our version includes the
following important modifications to DoPHOT: (1) utilizes a digitized
point source profile, rather than analytic, to best match the real
Spitzer data; (2) accepts floating point input FITS images and
computes output fluxes based on the Spitzer surface brightness units
in the BCD images; (3) accepts input masks to avoid using pixels
that have been declared bad for whatever reason; (4) includes a
multiframe mode that fits fluxes and positions from the entire stack
of BCD frames relevant to any input source position using the
un-smoothed instrumental PSF rather than that produced by the
mosaicking process.  The most significant unchanged aspect of
c2dphot/DoPHOT is the basic source extraction process.  In particular,
c2dphot starts at an upper flux level, finds and characterizes sources
above that level, and subtracts them from the image.  It then works
its way down in flux, typically by a factor of 2 step each time, doing
the same thing until it reaches the lower flux limit input by the
user. If c2dphot finds an object that is better fit by a two-axis
ellipsoid than the numerical point source profile, it will classify
the object as extended and produce estimates of the source size and
tilt of the ellipse.

Like any source extractor, there are many tunable parameters in
c2dphot to enable it to deal with a variety of problems or
characteristics in the data.  We discuss the most important here and
how we have tuned them for the sources found in this analysis.  The
most sensitive parameters for both accurate source extraction and
photometry are the sizes of the search box, fitting box, and aperture
photometry boxes.  For these data, we used search and aperture boxes of
7 pixels and a fitting box of 5 pixels, based on a number of tests
on simulated data sets as well as a variety of real data from Spitzer.
Other tunable parameters include thresholds for deciding whether a
source is extended, whether an initial detection is better fitted by a
tilted plane, and whether a source is so small that it is more likely
to be a previously undetected radiation hit.  These have all been
tuned to levels that appear appropriate from careful visual
inspections of subsets of the data.

The details of the c2dphot source extraction processing for the 24
\micron\ data involve running the source extractor first on the mosaic
image produced by MOPEX.  Then the multi-frame option is run with the
output list from the previous processing given as an input list for
flux, position, and shape refinement, but no new source searching.  As
mentioned above, this procedure insures that the source
characteristics are derived from the least-processed form of the data,
the BCD products. For MIPS 70 \micron\ data we have extracted sources
from the mosaics made with the filtered BCD images with c2dphot.

The calibration of source fluxes is done in the following way. For all
objects for which a reliable aperture flux could be determined and
which were well fitted by the nominal point source profile, the ratio
of the aperture flux to model flux was averaged. The fact that this
ratio is not unity is expected since the model fluxes are essentially
the product of a single number, the fitted peak value, multiplied by
the PSF area. Small errors and uncertainties in the PSF area will then
lead to errors in the total model flux. We then assume that this
average ratio of aperture flux to model flux applies for all point
sources and multiply by an additional correction factor of about a
factor of two at 24 \micron\ for the aperture used in c2dphot relative
to that for the absolute calibration used by the SSC (essentially a
traditional ``aperture correction''). This correction introduces
absolute flux uncertainties of up to 10\%.

Relative flux and position uncertainties in c2dphot are calculated in
a standard way from a numerical estimate of the Hessian matrix
(\citealt{pre97}, \citealt{siv96}). In particular, the matrix of
partial gradients of chi-squared is calculated numerically for
variations in the four model parameters (for point sources).  This is
done by fixing each parameter at levels slightly offset above and
below the best fit parameter and calculating the change in chi-squared
for all combinations of offset parameters.  The diagonal elements of
the square root of the inverse of this matrix then give an error
estimate, at least in the case where the errors are reasonably
behaved.  A random check of the off-diagonal elements has shown that
the only significant correlation between error estimates is that
expected between the derived sky level of the fit and the peak star
amplitude.  For extended extractions a similar procedure is part of
DoPHOT and c2dphot using analytic, rather than numerical estimates of
the derivatives.

The absolute uncertainty in the fluxes is at least 5\% at 24, 10\% at
70, and 20\% at 160 \micron\ as estimated by the MIPS instrument team
\citep{gor05b}. However, considering uncertainties introduced by the
``aperture correction'' in c2dphot, we estimate the total uncertainty
in these data to be closer to 10\% at 24 and 20\% at 70 and 160 \micron.

\subsubsection{Bandmerging}\label{bandmerge}

The 24 and 70 $\mu$m bands were processed differently in the
bandmerging stage of the c2d pipeline because the 70 $\mu$m band has
only one effective epoch of observations.  Our observations were
planned so that our second epoch would fill in the gaps in the
coverage at 70 $\mu$m caused by the unusable half of the array.

The final list of sources and fluxes was produced in three
steps. First, the source extractions for 24 $\mu$m in the two epochs
are compared.  A source detection in one epoch that has only a single
match within 2\farcs0 in the other epoch and no other matches in its
own epoch is considered a ``good'' source and the two detections are
averaged. Detections that do not have any match within 2\farcs0 in
either epoch or detections with more than one match in an epoch are
suspect sources and these are checked to ensure no valid sources are
missed.  The candidate sources are then filtered to include only those
sources with fluxes greater than or equal to 10 times the flux
uncertainty (10-$\sigma$). Asteroids, which are bright at 24 $\mu$m,
shift positions between the two epochs.  As a result, they are
excluded in this step because an asteroid detected in one epoch will
not have a corresponding detection within 2\farcs0 in the second
epoch. As expected, due to Cha II's ecliptic latitude of $-$65\degree,
no asteroids were detected at 24 \micron\ in the data.  Asteroids are
not bright enough at 70 \micron\ to be detected.

In step two, the filtered source extractions for 24 $\mu$m are
compared with the source extractions made from the combined-epochs
images.  A good source detection has only a single match within
2\farcs0 in both the filtered list and the combined epochs list.  For
these sources, the position and flux information from the combined
epochs detection is substituted as the best value for the source.  As
in step 1, sources not detected within 2\farcs0 in both the filtered
list and the combined-epochs list or sources containing multiple
detections within 2\farcs0 in either list are examined and saved if
merited.  The final list of good sources at 24 $\mu$m is again filtered
to only include those with fluxes greater than or equal to
10-$\sigma$.  Because we only have one epoch of observations at 70
$\mu$m, only the sigma clipping was performed in this step for the 70
$\mu$m data.

The 24 and 70 $\mu$m bands were merged in step three.  The source
extraction program produced many artificial detections near the ends
of the MIPS scan legs at 70 \micron.  Most of these false detections
were eliminated from the source list by the 10-$\sigma$ cut and by
removing sources with aperture fluxes less than or equal to zero. The
final 70 \micron\ source list was then bandmerged with the 24 \micron\
source list.

\subsection{SIMBA Data}

The SIMBA 1 mm mosaics were reduced and combined using the
MOPSIC\footnote{MOPSIC is a software package for infrared, millimeter,
and radio data reduction and analysis developed and constantly
upgraded by R.\ Zylka.} software package. The noise in the maps forces
us to subtract correlated skynoise and baselines of order 2, which
both filter out emission on large scales. The correlated skynoise was
evaluated as a weighted mean of the channels spaced by more than
50\arcsec\ from the channel to be corrected. Thus the response to
extended sources (i.e., with a diameter larger than the
``correlation'' radius of 50\arcsec) is damped by the skynoise
filter. Such sources are not quantitatively preserved. Compact (i.e.,
diameter less than 50\arcsec) and unresolved sources (i.e., diameter
less than a beam) are not significantly affected by the filter. In
total, 7 noisy maps and several subscans with spikes and drifts were
not included in the mosaic in order to improve the noise level and to
minimize artifacts. The final on-source integration time is therefore
$8.4 ~ \rm hours$.

We apply a reduction scheme with three iteration steps, which reduces
the filtering of extended emission by skynoise subtraction. After the
first step, regions with significant emission are identified and
included in a source model. In the second step, the source model is
first subtracted from the deconvolved raw data before skynoise and
baseline removal, and then again added to the data before coadding.
This yields a map in which the extended emission is better preserved
from spatial filtering by the skynoise filter, while the impact of
baseline removal is not mitigated. The improved map is then used to
obtain an improved source model, which then is used in a third and
final iteration step to derive the final map. We correct for pointing
uncertainties by shifting maps by the offset between the SIMBA and
MIPS positions of IRAS 12496-7650 observed for the respective epoch
(i.e., by 7\arcsec\ to 16\arcsec).

\section{Results}

\subsection{MIPS Source Counts}

1532 sources at 24 \micron\ and 41 sources at 70 \micron\ were
identified with fluxes greater than 10-$\sigma$.  Normal source
extraction techniques were not possible on the 160 \micron\ data
because the emission is so extended and the coverage was
incomplete. However, several bright peaks are seen in the map, and 160
\micron\ aperture fluxes are presented for two sources discussed in
Section \ref{int}.  In the mosaics, the median background and rms
noise values in the off-cloud regions where structured dust emission
is low are 16 and 0.1 MJy ster$^{-1}$ at 24 \micron, 6 and 0.8 MJy
ster$^{-1}$ at 70 \micron, and 30 and 4.5 MJy ster$^{-1}$ at 160
\micron.

\begin{figure}
\plotone{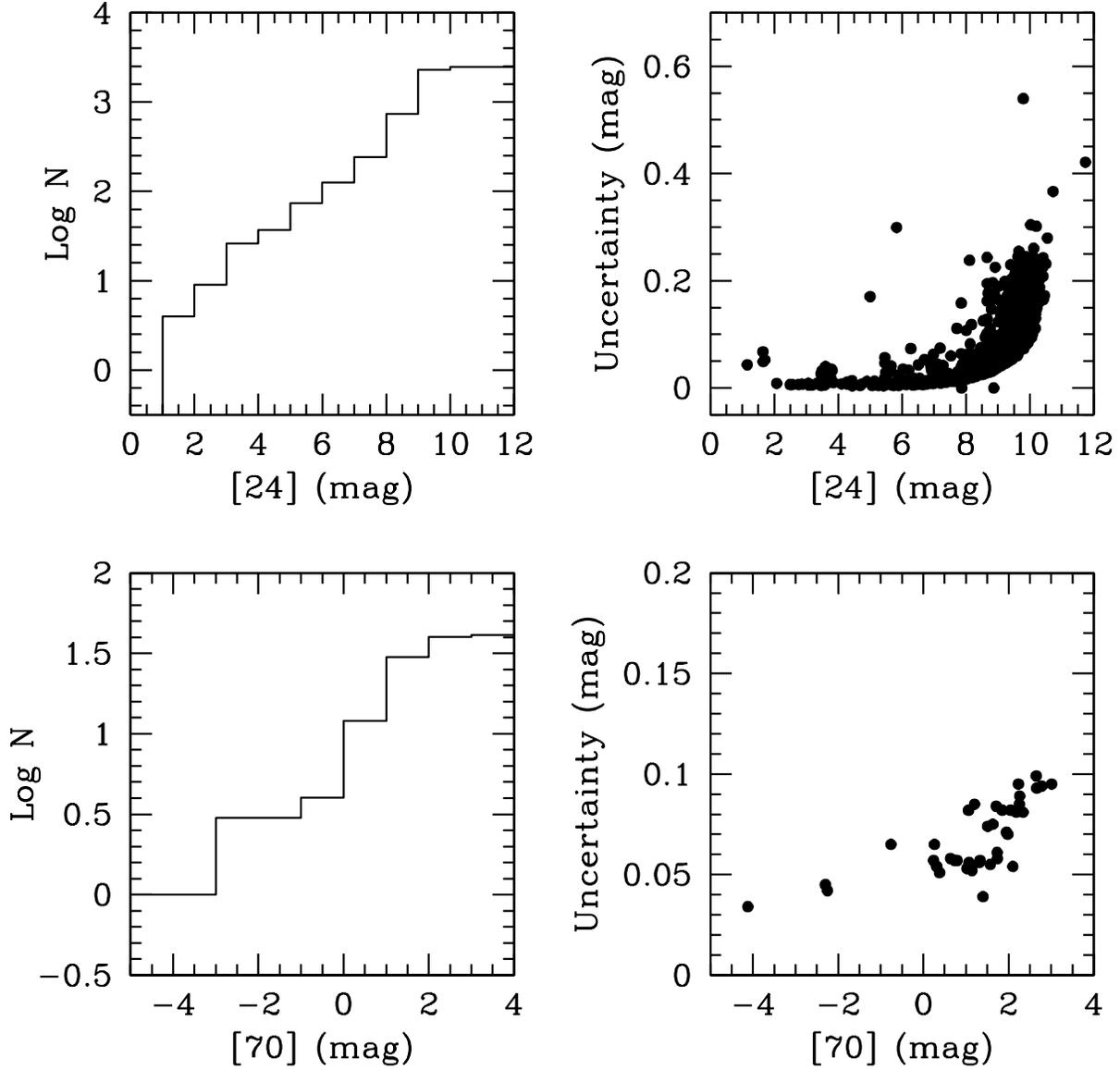} \figcaption{\label{comp} Left panels: Log of
the cumulative number of counts for 24 (top) and 70 \micron\ (bottom)
sources per magnitude bin.  Right panels: Uncertainty of extracted
magnitude versus magnitude for 24 (top) and 70 \micron\ (bottom)
sources. The 24 \micron\ counts included all sources extracted from
the Cha II map (2481 sources).  The 70 \micron\ counts are only
sources with fluxes greater than 10-$\sigma$ and identified as real by
eye (41 sources). The estimated completeness from these plots at 24
and 70 \micron\ is 10 mag (0.7 mJy) and 3 mag (50 mJy) respectively.}
\end{figure}

To determine approximate completeness limits, we plot two parameters
versus source magnitude: cumulative source counts and extracted
uncertainty.  Figure \ref{comp} shows the cumulative source counts for
24 and 70 \micron\ in Cha II.  The 24 \micron\ counts include all
sources extracted from the mosaic with no sigma cut (2481
sources). The 70 \micron\ cumulative source counts only include
greater than 10-$\sigma$ sources verified by eye to be real (41
sources) since the source extraction produced many artificial
detections at the end of the scan legs in this band. The 24 and 70
\micron\ magnitudes were calculated using zero point fluxes derived
from a blackbody corresponding to Vega (Table \ref{zero}). The source
counts flatten at the faintest magnitudes, indicating completeness to
about 10 mag (0.7 mJy) at 24 \micron\ and 3 mag (50 mJy) at 70
\micron. Plots of magnitude uncertainty versus magnitude for Cha II,
Figure \ref{comp}, show that the uncertainty reaches 0.1 magnitude at
close to the same values as the turnover in source counts, 10 mag at
24 \micron\ and 3 mag at 70 \micron. Experience with other large
surveys suggests that 0.1 magnitude uncertainty is usually about where
the 90\% completeness level lies. The 24 \micron\ flux limit of 0.7
mJy is consistent with the 3-$\sigma$ sensitivity (0.83 mJy for 24
seconds integration) predicted for the c2d program by
\citet{eva03}. However, the 70 \micron\ limit of 50 mJy is much higher
than the 10-$\sigma$ sensitivity predicted prior to launch (17 mJy in
24 seconds; \citealt{eva03}) partially due to the loss of half of the
array and resultant cut in the integration time. The observed
completeness limits are about 5 and 6 times the SSC's expected
post-launch 1-$\sigma$ sensitivities with low background and
appropriate integration times at 24 and 70 \micron\ respectively
(Spitzer Observer's Manual 2004).

\begin{figure}
\plotone{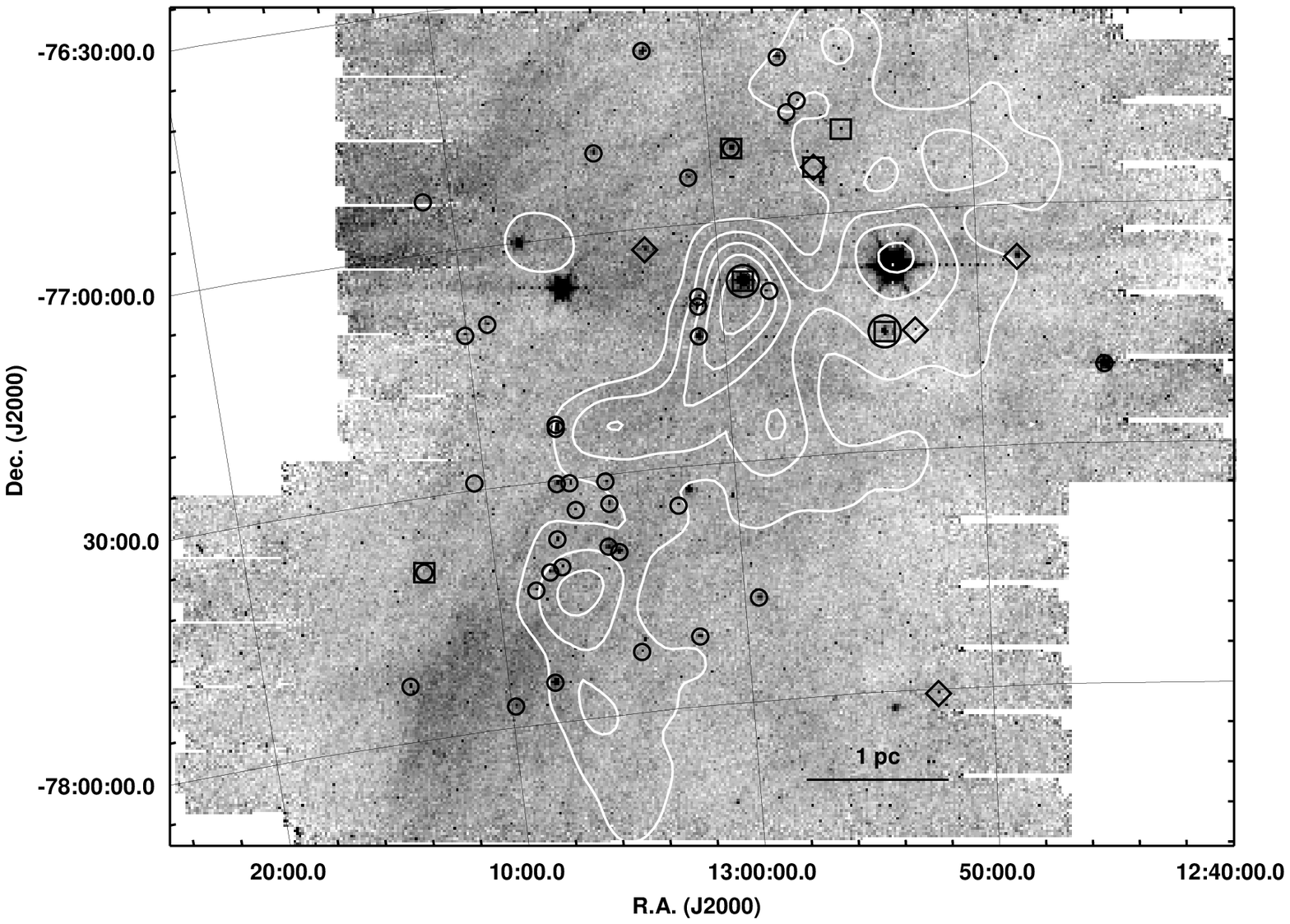} \figcaption{\label{av_yso} MIPS 24 \micron\ map of
the Cha II molecular cloud with A$_V$ contours from Vuong et al.\
(2001) in white. The contours are A$_V$ = 2 to 10 mag by 2 mag. The
circles and diamonds indicate the positions of the sources listed in
Table \ref{yso}. The different symbols identify the class of
the object according to the $\alpha_{24/K}$ given in the table. Class
I are large circles, Class II are small circles, and Class III are
diamonds. The squares mark the positions of the selected sources
discussed in Section \ref{int}. The three brightest sources from East
to West are IRAS 13022-7650 (DL Cha, variable star), IRAS 12553 (Class
I), and IRAS 12496 (DK Cha, Ae star). Latent images are seen as a line
of dots to the East and West of these sources. The bright source to
the NE of IRAS 13022 at $-$77\degree\ is IRAS 13036 (Class 0/I) in the
molecular core BHR 86.}
\end{figure}

\begin{figure}
\plotone{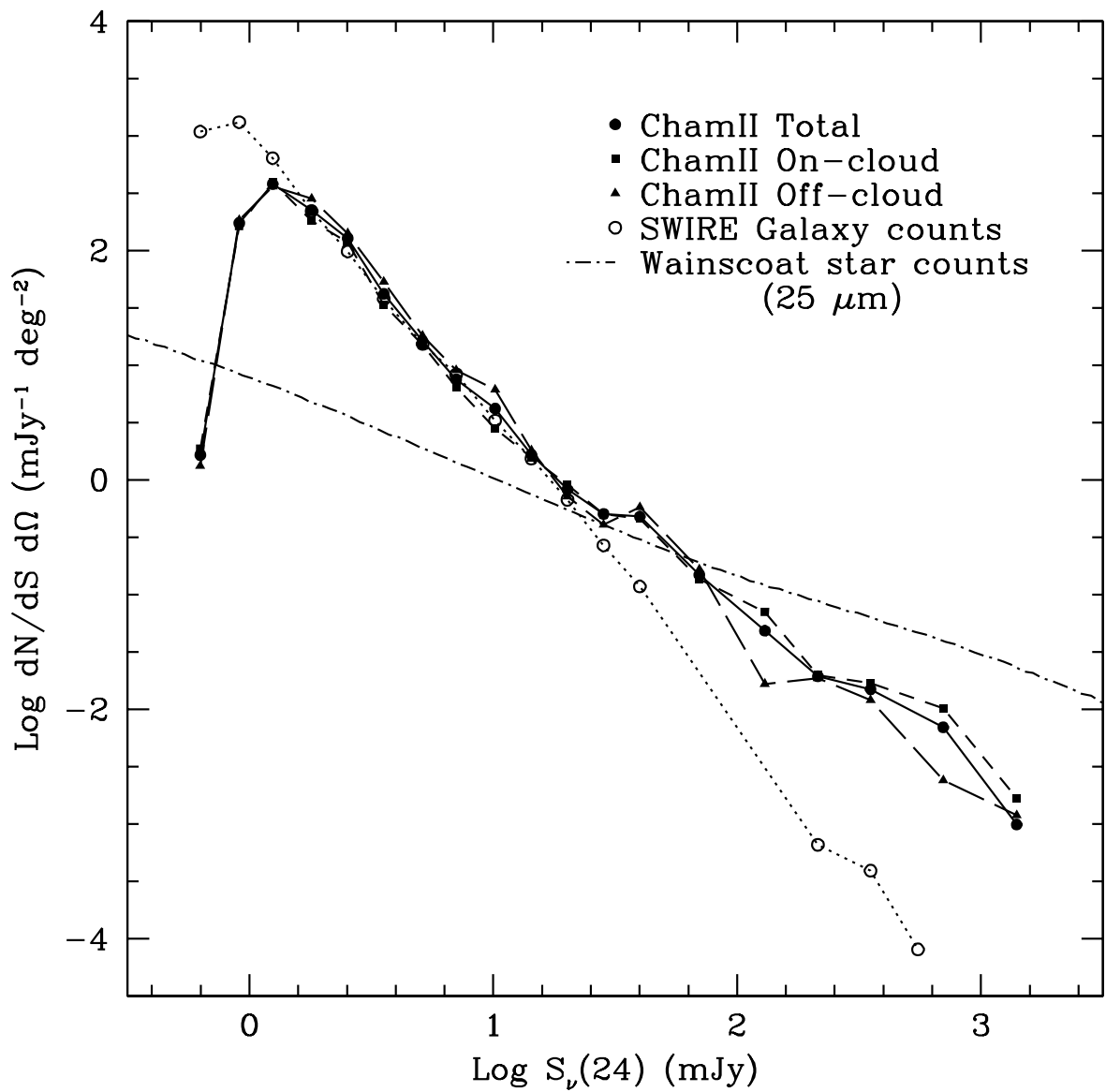} \figcaption{\label{dnds} Source counts in the Cha II
MIPS field. On-cloud counts refer to the area of the map with A$_V >
1$. Off-cloud counts are where A$_V < 1$. The SWIRE galaxy counts are
from Marleau et al.\ (2004). The dot-dash line is model star counts at
25 \micron\ for the latitude, longitude, and distance of Cha II with
A$_V$ = 2 from Wainscoat et al.\ (1992). The source counts are consistent
with the galaxy counts for 24 \micron\ fluxes less than 20 mJy. However, 
brighter sources are likely Galactic.}
\end{figure}

Figure \ref{dnds} shows the 24 \micron\ source counts per square
degree versus flux ($dN/dSd\Omega$ vs.\ $S_{\nu}(24)$) for the total
1532 sources in the complete mosaic of Cha II.  The sample is flux
limited as shown in Figure \ref{dnds} by the steep drop in
$dN/dSd\Omega$ for $S_{\nu}(24) < 0.7$ mJy. The complete field can be
divided into two regions: on-cloud and off-cloud. The on-cloud field
is defined by an area that encloses the \citet{cam99} A$_V$ = 1
contour and includes 825 24 \micron\ sources. The remainder of the
field is considered off-cloud.  Figure \ref{dnds} compares
$dN/dSd\Omega$ of the total field (on-cloud and off-cloud samples)
with galaxy counts from the SWIRE Spitzer Legacy project
(\citealt{lon03}; \citealt{mar04}) and model star counts for the
position of Cha II at 25 \micron\ from \citet{wai92}. The Wainscoat
star counts were computed using a C-version of the model provided by
J. Carpenter (2001, private communication). This version of the model
allows for the inclusion of extinction in calculating the source
counts. The A$_V$ was set to 2 magnitudes as an approximate average
between off-cloud, where A$_V$ $<$ 1, and on-cloud regions, where
A$_V$ = 1 to 6 with a few patches of A$_V$ $>$ 6 (Fig.\ \ref{av_yso}).

For $S_{\nu}(24) < 20$ mJy, the Cha II source counts are consistent
with the SWIRE galaxy counts.  For fluxes greater than about 20 mJy,
the Cha II and model star counts exceed the galaxy counts.  Therefore,
most of the bright 24 \micron\ sources are likely associated with the
Cha II cloud or are Galactic, while most of the fainter sources are
probably galaxies. The \citet{wai92} model star counts also exceed the
Cha II source counts for $S_{\nu}(24) > 70$ mJy. Even though the total
source and model counts are small in these flux bins, we assume not
all of the Cha II sources are background stars. The discrepancy
between the Wainscoat model and the data appears real and indicates
that the model overestimates the star counts toward Cha II.
Increasing the A$_V$ in the Wainscoat model reduces the
inconsistency. However, an A$_V$ of 20 magnitudes is needed for the
model counts to be approximately equal to the total source counts for
bright sources. This large extinction is only seen for a few isolated
positions in the cloud so is not the best choice for the model.  For
24 \micron\ fluxes greater than 70 mJy, there is an excess of on-cloud
sources over off-cloud sources (21 on-cloud vs.\ 7
off-cloud). Although this excess is statistically small, the numbers
are representative of the locations of previously known YSOs in Cha
II. There are a small number of T Tauri-type stars associated with Cha
II in the off-cloud (A$_V <$ 1) area. Figure \ref{av_yso} shows the
Spitzer 24 \micron\ map of Cha II with A$_V$ contours and known and
candidate YSOs marked. This figure shows that most of the YSOs in Cha
II are on the eastern edge of the cloud rather than in areas of high
extinction.

\subsection{MIPS--2MASS Color-Magnitude Diagrams} \label{colors} 
			   	
\begin{figure}
\plotone{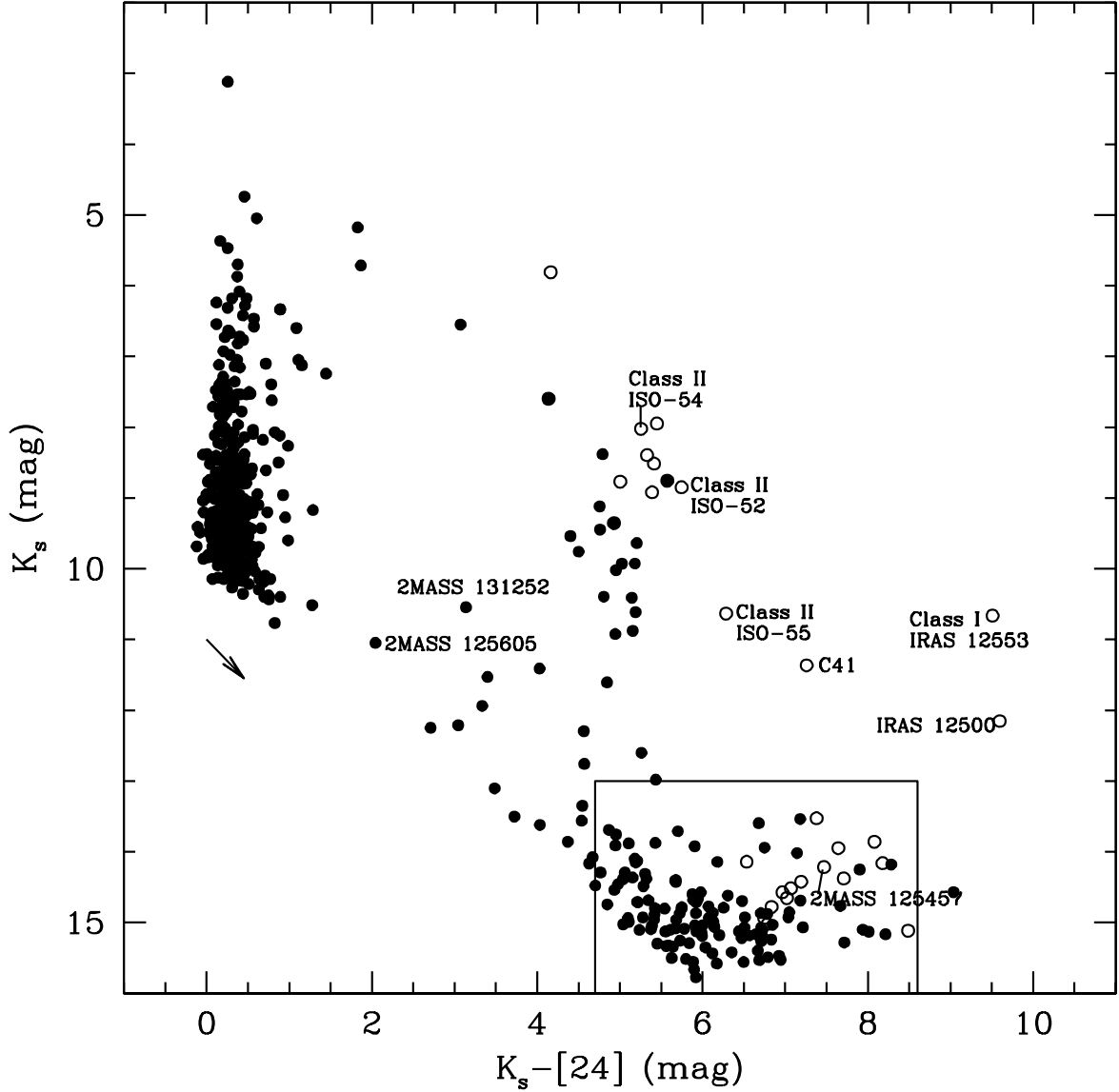} \figcaption{\label{k24} Color-magnitude diagram for
the 2MASS K$_s$ band and the MIPS 24 \micron\ band sources in Cha
II. 626 (41\%) of the 24 \micron\ MIPS sources in Cha II have K$_s$
detections. Sources with 70 \micron\ detections are open circles. The
box denotes where the SWIRE galaxies would fall on this plot. Known
Class I and II sources from Persi et al.\ (2003) and sources described
in Section \ref{int} are labeled. The arrow is the reddening vector
for A$_V$ = 5 mag using the \citet{wei01} dust model with $R_V$ =
5.5. Sources with K$_s$ $<$ 13 mag and (K$_s$ $-$ [24]) $>$ 1 are
considered YSO candidates.}
\end{figure}

The 24 \micron\ source list was bandmerged with the 2MASS catalog
\citep{cut03} for matches within  2\farcs0.  We identified 626 sources
as having both a 24 \micron\ and K$_s$ (2.159 \micron) detection. The
magnitude limit of the 2MASS catalog is K$_s \sim$ 15 mag. These
sources are plotted in a K$_s$ magnitude versus (K$_s-$[24]) color
diagram in Figure \ref{k24}. The diagonal cutoff at lower left of the
figure indicates the 24 \micron\ flux limit of around 0.7 mJy.  Two
large clusters are seen in the plot. The first clump at (K$_s-$[24]) =
0 contains main sequence stars. The second cluster at K$_s$ $>$ 13 mag
and (K$_s-$[24]) $>$ 4 mag probably consists mainly of galaxies.
Using data from the SWIRE extragalactic survey with Spitzer
(\citealt{lon03}; \citealt{sur04}), we predict where galaxies are
likely to fall in the K$_s$ versus (K$_s-$[24]) color-magnitude
space. Most of the SWIRE galaxies fall between (K$_s-$[24]) = 4.5 and
8.5 mag with K$_s$ $>$ 13 mag, indicated by a box in Figure
\ref{k24}. A few SWIRE objects trail out between (K$_s-$[24]) = 3 and
4.5 magnitudes, all of those with K$_s$ $>$ 12 mag, indicating that
some of the sources in the plot that are faint at K$_s$, but with
(K$_s-$[24]) $<$ 4.5 might also be galaxies. However, 5 of the 11
sources with 11 $<$ K$_s$ $<$ 13 and (K$_s-$[24]) $<$ 5.5 mag were
identified as YSO candidates by \citet{vuo01} or
\citet{per03}. Therefore, that region of color-magnitude space is
ambiguous.

Reddened sources with relatively bright K$_s$ magnitudes between the
star and galaxy clusters in the K$_s$ vs.\ (K$_s-$[24]) plot (Fig.\
\ref{k24}) are potential YSOs associated with Cha II.  Class II
sources fall in the middle of the plot with bright K$_s$ magnitudes
and red (K$_s-$[24]) colors. The Class I sources IRAS 12553-7651
(ISO-ChaII-28, \citealt{per03}) and IRAS 12500-7658 \citep{che97} are
outliers with the reddest (K$_s-$[24]) colors. In order to find
potential YSOs, we identified sources with K$_s <$ 13 mag and
(K$_s-$[24]) $>$ 1 mag. These criteria eliminate galaxies and stars,
but also generally exclude Class III sources, which have similar
colors to stars. Forty-nine sources fit these criteria and were
cross-checked with known YSOs from the literature. Forty were
previously identified as YSOs or YSO candidates (\citealt{gau92};
\citealt{pru92}; \citealt{che97}; \citealt{vuo01};
\citealt{per03}). Five of the sources have SEDs consistent with highly
extincted background stars. Two sources are possible YSOs but are very
faint at K$_s$ ($>$ 12 mag) and cannot be distinguished from galaxies
with these data. However, the SEDs of 2 sources (2MASS
12560549-7654106 and 2MASS 13125238-7739182) indicate they have a 24
\micron\ excess and are likely stars with disks. These sources are
described further in Section \ref{int}. The 40 previously identified
and 4 new YSO candidates selected from the color-magnitude diagram are
listed in Table \ref{yso}. This list of YSO candidates includes 6
identified by Persi et al.\ (2003, Table 3 \& 4) with ISO, including
one of their new candidate YSOs ISO-ChaII-13. Our YSO criteria were
unable to identify the Class III sources from \citet{per03}, and we
did not detect the 3 other new candidates from their Table 4 with
MIPS.  Table \ref{yso} also includes 9 low mass T Tauri or young brown
dwarf candidates identified by \citet{vuo01}.

Despite the improved sensitivity of MIPS over previous infrared
instruments, the small number of new YSO candidates found with MIPS
and 2MASS data in Figure \ref{k24} indicates that previous studies
(e.g., \citealt{sch77}; \citealt{gau92}; \citealt{pru92};
\citealt{per03}) achieved a nearly complete census of the YSOs in Cha
II. However, a later paper will combine the MIPS and IRAC data of Cha
II. The IRAC data may allow the identification of more new low
luminosity YSOs among the sources that cannot be distinguished from
galaxies with MIPS and 2MASS data alone.

\begin{figure}
\plotone{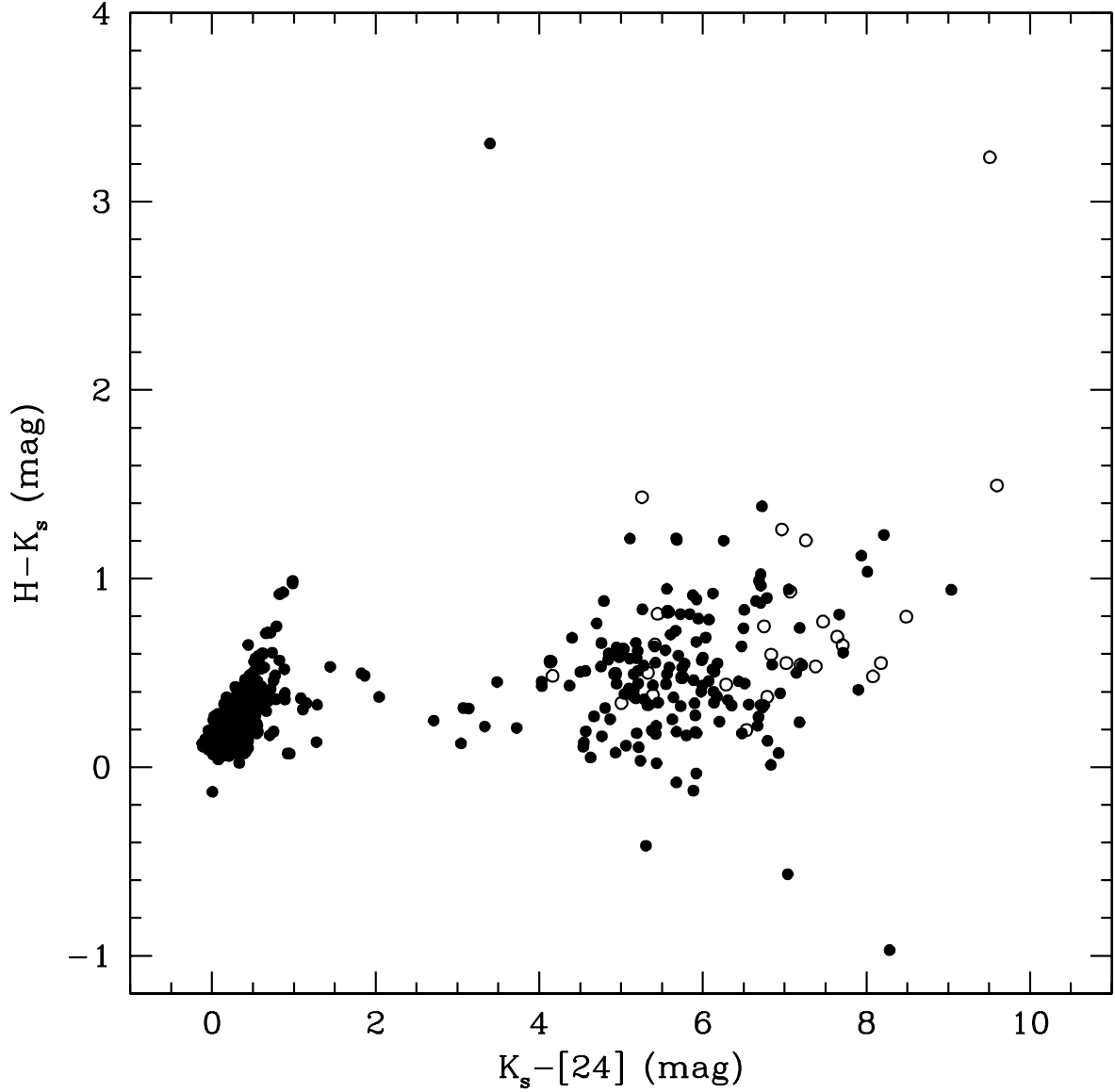} \figcaption{\label{hk24} Color-color diagram
comparing 2MASS (H$-$K$_s$) vs. (K$_s-$[24]). Sources with 70 \micron\
detections are open circles. Sources with (K$_s-$[24]) $>$ 3 mag cannot
be distinguished from galaxies in this color-color space.}
\end{figure}

Figure \ref{hk24} plots (H$-$K$_s$) versus (K$_s-$[24]). In this
color-color space, the SWIRE (\citealt{lon03}; \citealt{sur04})
galaxies fill nearly the same area as the data with (K$_s-$[24]) $>$ 3
mag. Therefore, it is very difficult to use this color-color space to
identify potential YSOs. A future paper on IRAC observations of Cha II
by the c2d team (Porras et al.\ 2005, in preparation) explores
different color-color spaces including the IRAC bands for identifying
YSOs.

\begin{figure}
\plotone{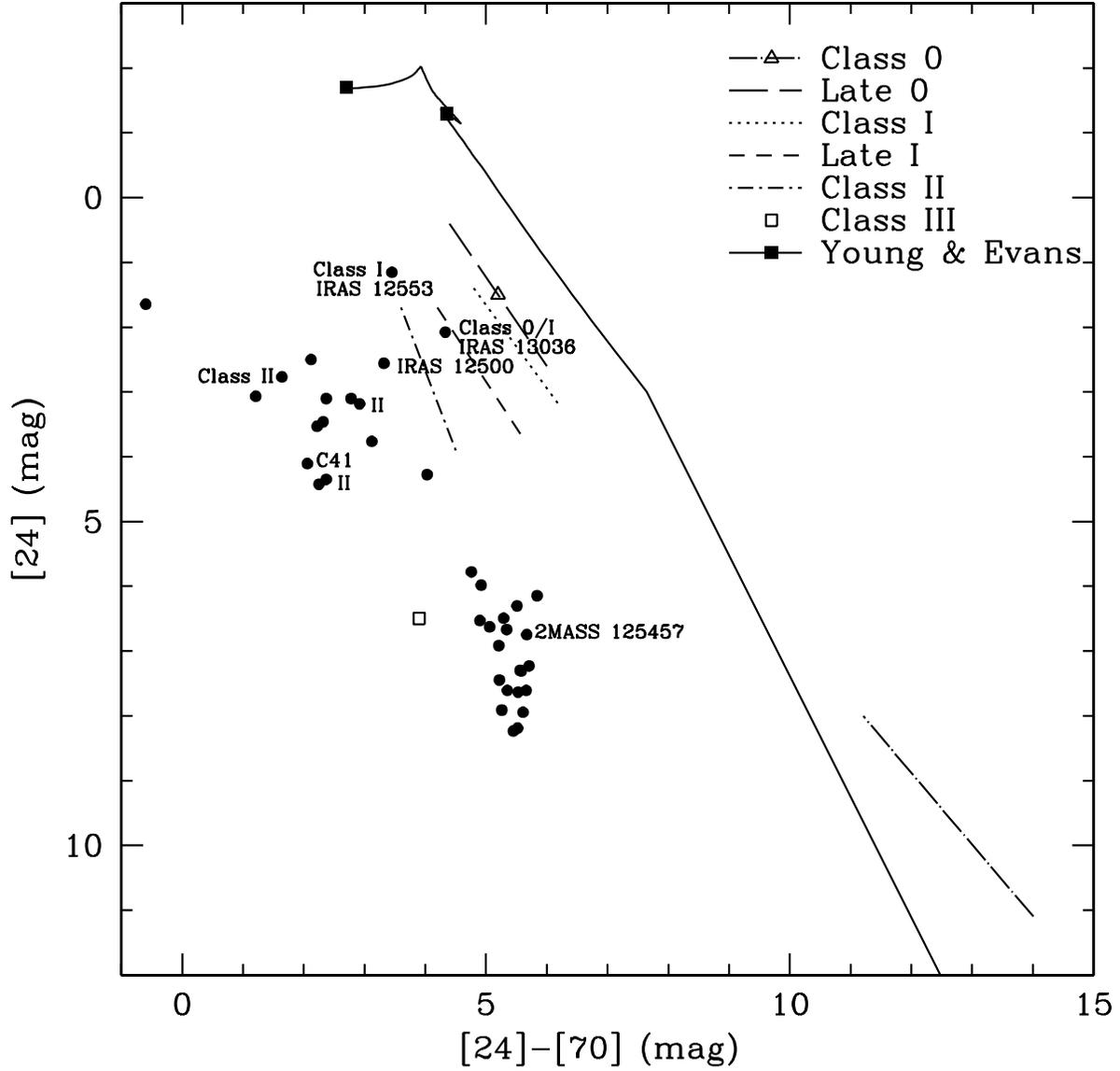} \figcaption{\label{24-70} Color-magnitude plot of
[24] vs.\ ([24]$-$[70]). Model predictions for a low mass protostar
from Whitney et al.\ (2003; Fig.\ 8a) are plotted as lines for Class 0
-- II and as a square for Class III. The lines span differences in
inclination angle with edge-on at the faint end and are for an
aperture size of 1000 AU. The solid line is an evolutionary track for
a 1 \msun\ protostar from Young \& Evans (2005). The solid squares
indicate transition points in the Young \& Evans model between Class
0/I and Class I/II. Known Class I and II sources \citep{per03} and 
selected sources described in the text are labeled and are generally
bluer than predicted by the models.}
\end{figure}

A MIPS 24 and 70 \micron\ color-magnitude diagram is shown in Figure
\ref{24-70}. Thirty-eight Cha II sources with detections at both 24
and 70 \micron\ are distributed in two clumps on the color-magnitude
plot, the first with [24] $<$ 5 mag and the second with [24] $>$ 6
mag.  Known Class I and II sources from \citet{per03} as well as some
other interesting objects are labeled on the plot. The Class II
sources fall within the main cluster of sources near ([24]$-$[70]) = 2
mag. Three Class 0/I sources, IRAS 13036-7644 (in an isolated core to
the East of the main cloud, BHR86; \citealt{bou95}; \citealt{mar97}),
IRAS 12553 (\citealt{per03}) and IRAS 12500 \citep{che97}, fall to the
right of the group with redder ([24]$-$[70]) colors.  All of the
sources in the second group, with [24] $>$ 6 mag, are very faint ($>$
13 mag) or not detected at K$_s$. Since the SWIRE galaxies all have
similarly faint K$_s$ magnitudes, the second group cannot be
distinguished from galaxies. The separation between the groups in
Figure \ref{24-70} also suggests that objects brighter than 6 mag at
24 \micron\ are likely within the Galaxy.

Models of 1 \lsun\ Class 0 to Class III protostars from \citet[see
their Fig.\ 8a]{whi03} and an evolutionary track of a 1 \msun\
protostar from 4000 to 210,000 years (Class 0 and I stages)
\citep[their Fig.\ 20]{you05} are also plotted on Figure
\ref{24-70}. Each \citet{whi03} line on the plot represents a range of
inclination angles for a model of a particular class object observed
with a 1000 AU radius aperture. This aperture would be roughly
equivalent to a diameter of 4 24-\micron\ pixels or 2 70-\micron\
pixels for the distance of Cha II. \citet{you05} also give
evolutionary tracks for 0.3 and 3 \msun\ protostars, but these do not
vary significantly from the 1 \msun\ track, except that the 0.3 \msun\
track drops down to [24] $\approx$ 0 mag for ([24]$-$[70]) $<$ 4
mag. Most of the Cha II sources with bright 24 \micron\ magnitudes
have bluer ([24]$-$[70]) colors than predicted either by \citet{whi03}
or by the evolutionary track of \citet{you05}. However, the Young \&
Evans (2005) evolutionary track and most of the \citet{whi03} models
are for younger protostars (Class 0 -- I), while Cha II has only a few
Class I sources. The Cha II sample is dominated by more evolved Class
II objects. The second group of sources, with [24] $>$ 6 mag, falls to
the right (greater ([24]$-$[70]) colors) of the Whitney et al.\ (2003)
prediction for Class III sources, but these sources cannot be
distinguished from galaxies with these data alone.

\begin{figure}
\plotone{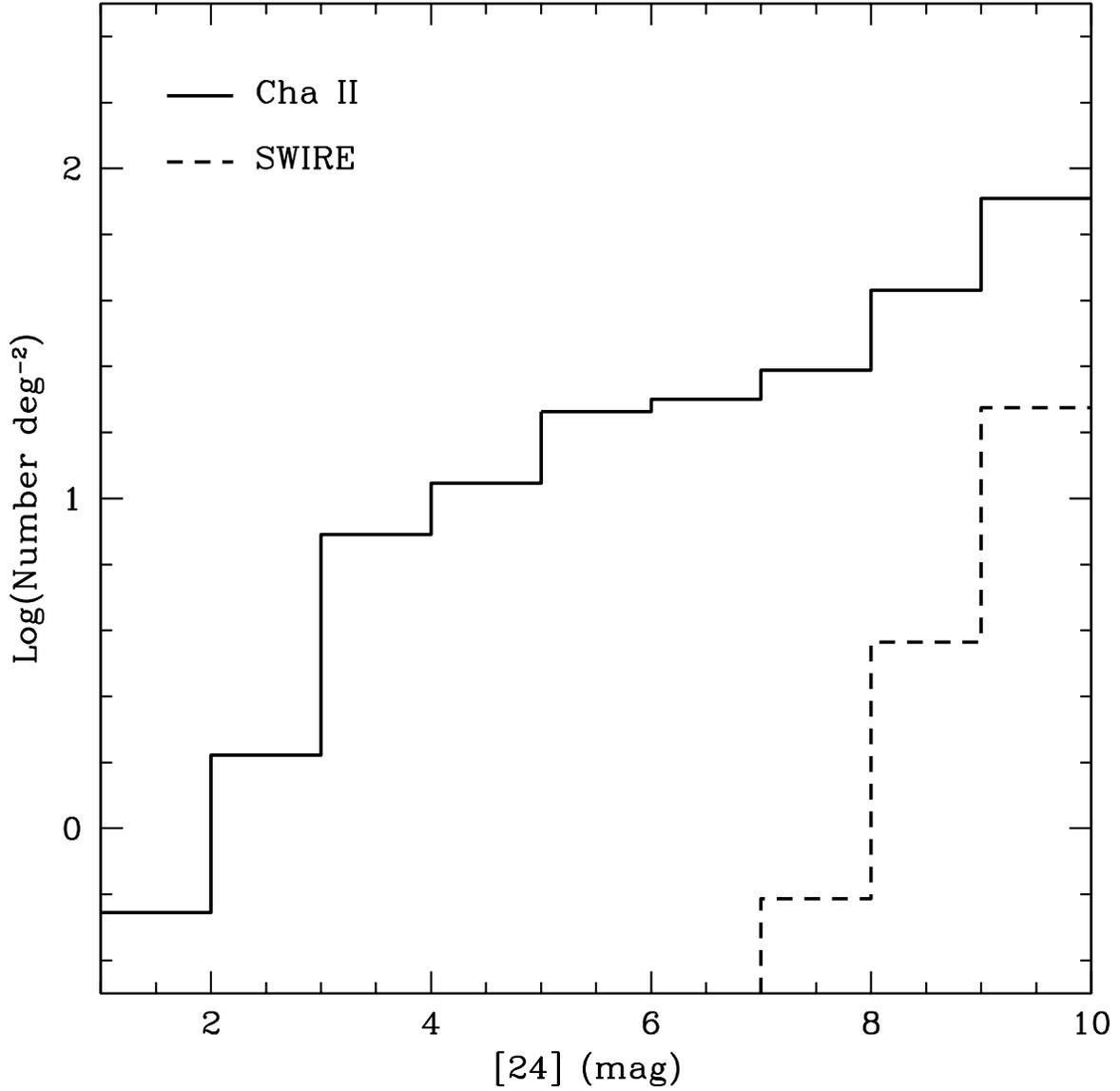}\figcaption{\label{cum} Histogram of the log of
the cumulative 24 \micron\ source counts per square degree for Cha II
(solid line) and the SWIRE extragalactic survey (dashed line) for
sources with $1<(K_s-[24])<6$ mag. Sources brighter than 7 mag at 24 \micron\ 
are not likely to be extragalactic.}
\end{figure}

Figure \ref{cum} compares the cumulative 24 \micron\ source counts per
square degree found in Cha II with the SWIRE survey (\citealt{lon03};
\citealt{sur04}). The histogram is limited to sources with
$1<(K_s-[24])<6$ to eliminate background stars. The plot confirms that
objects that are bright at 24 \micron\ are not likely to be
extragalactic. The Cha II source counts dominate the SWIRE counts for
all 24 \micron\ sources brighter than 8 -- 9 mag, and essentially no
galaxies are expected to be detected with magnitudes less than 7.

\subsection{MIPS--2MASS Spectral Index}

\begin{figure}
\plotone{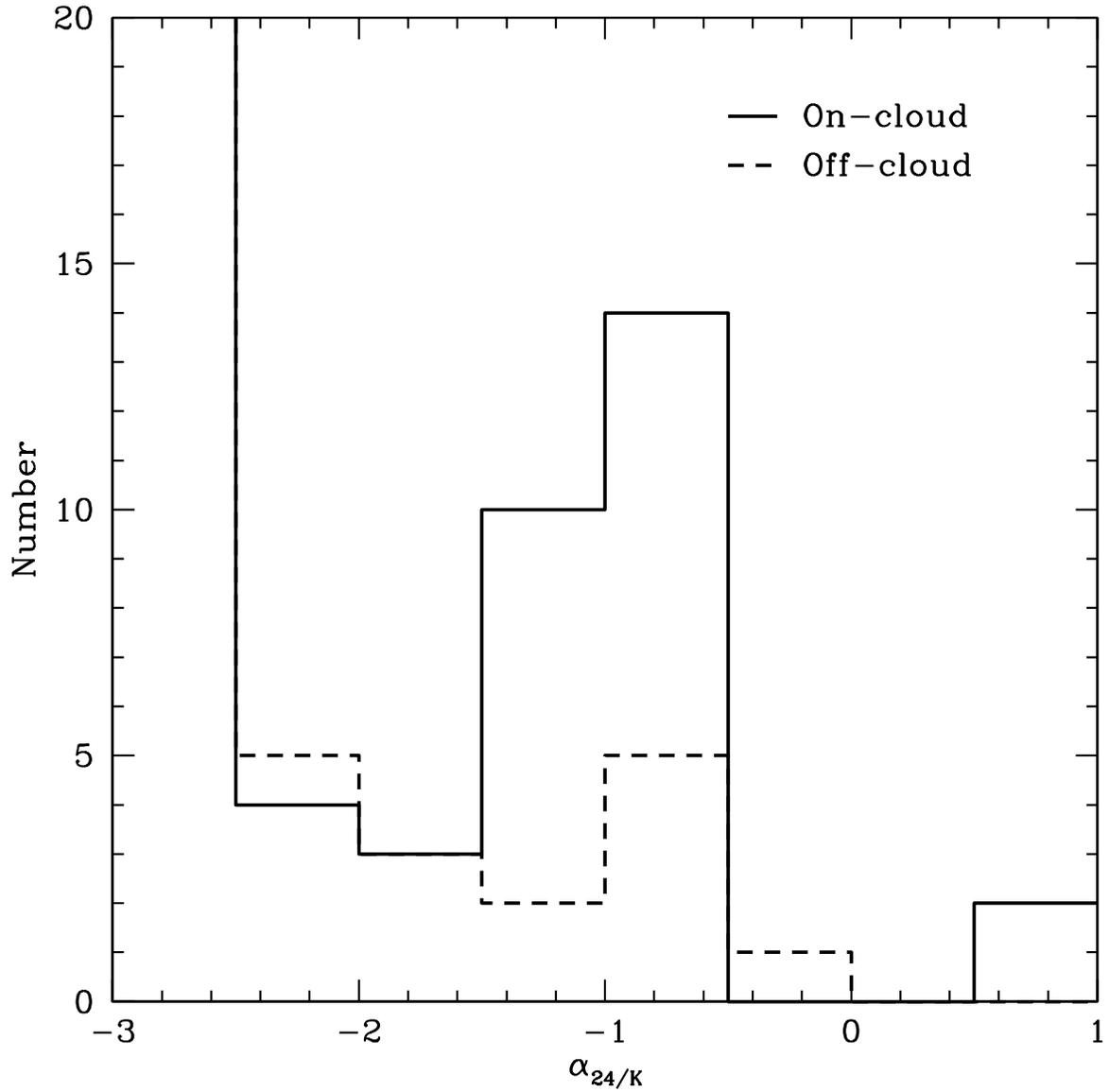} \figcaption{\label{index} Histogram of
spectral indices ($\alpha_{24/K}$; Lada 1987) of sources detected at
K$_s$ and 24 \micron\ with K$_s <$ 13 mag. On-cloud sources as defined
in the text are shown as the solid histogram. Off-cloud sources are
shown in the dashed histogram. The $\alpha_{24/K} <$ $-$2.5 bin
continues up to 233 on-cloud sources and 194 off-cloud sources. 
Sources with $\alpha_{24/K} >$ $-$2.5 are potential YSOs.}
\end{figure}

The spectral index for each source with 24 \micron\ and K$_s$ detections
was computed according to the equation:
\begin{equation}
\alpha_{24/K} = \frac{{\rm
log}((24S_{\lambda}(24))/(2.159S_{\lambda}(K_s)))}{{\rm
log}(24/2.159)}.
\end{equation}
A histogram of the spectral indices is shown in Figure \ref{index}.
The figure does not include sources with K$_s >$ 13 mag in order to
eliminate as much confusion from galaxies as possible.  The figure
shows the histograms for on-cloud and off-cloud sources as defined
above for the plot of source counts (Fig.\ \ref{dnds}). Most of the
sources (90\%) have a spectral index less than $-$2.5, with an average
of $-$2.8, which is consistent with that of main sequence stars.  For
clarity, the first bin of the histogram, $\alpha_{24/K} < -2.5$, is
not shown to its full extent. There are 233 on-cloud sources and 194
off-cloud sources in the first bin. \citet{lad87} defined a
near-infrared spectral index for wavelengths from 2 to 20 \micron\ and
used it to define boundaries for the Class system of YSO
classification. According to \citet{lad87}, a spectral index greater
than zero is a Class I source, $-2 < \alpha \le 0$ is a Class II, and
$-3 < \alpha < -2$ is a Class III. Forty-nine (10\%) of the sources
have $-2.5 < \alpha_{24/K} < 1$ indicating that these sources are
possible Class I -- III YSOs according to the \citet{lad87}
classification. All 49 are the same sources identified as potential
YSOs in the K$_s$ vs.\ (K$_s-$[24]) plot (Fig.\ \ref{k24}) as
described above. Two sources have $\alpha_{24/K} > 0$, IRAS 12500
and IRAS 12553, both previously classified as Class I protostars (see
Section \ref{int}). Both the on-cloud and off-cloud regions contain
sources with $\alpha_{24/K} > -2.5$. This is not surprising since
there are known T Tauri stars outside of the A$_V$ = 1 contour used to
define the on-cloud region (Fig.\ \ref{av_yso}). The spectral indices
of the 44 YSO candidates identified in the K$_s$ vs.\ (K$_s-$[24])
plot are listed in Table \ref{yso}. Figure \ref{av_yso} shows the
location of the sources in Table \ref{yso} with different symbols
indicating their spectral index classification.

\subsection{MIPS Extended Emission}\label{environ}
	
\begin{figure}
\plotone{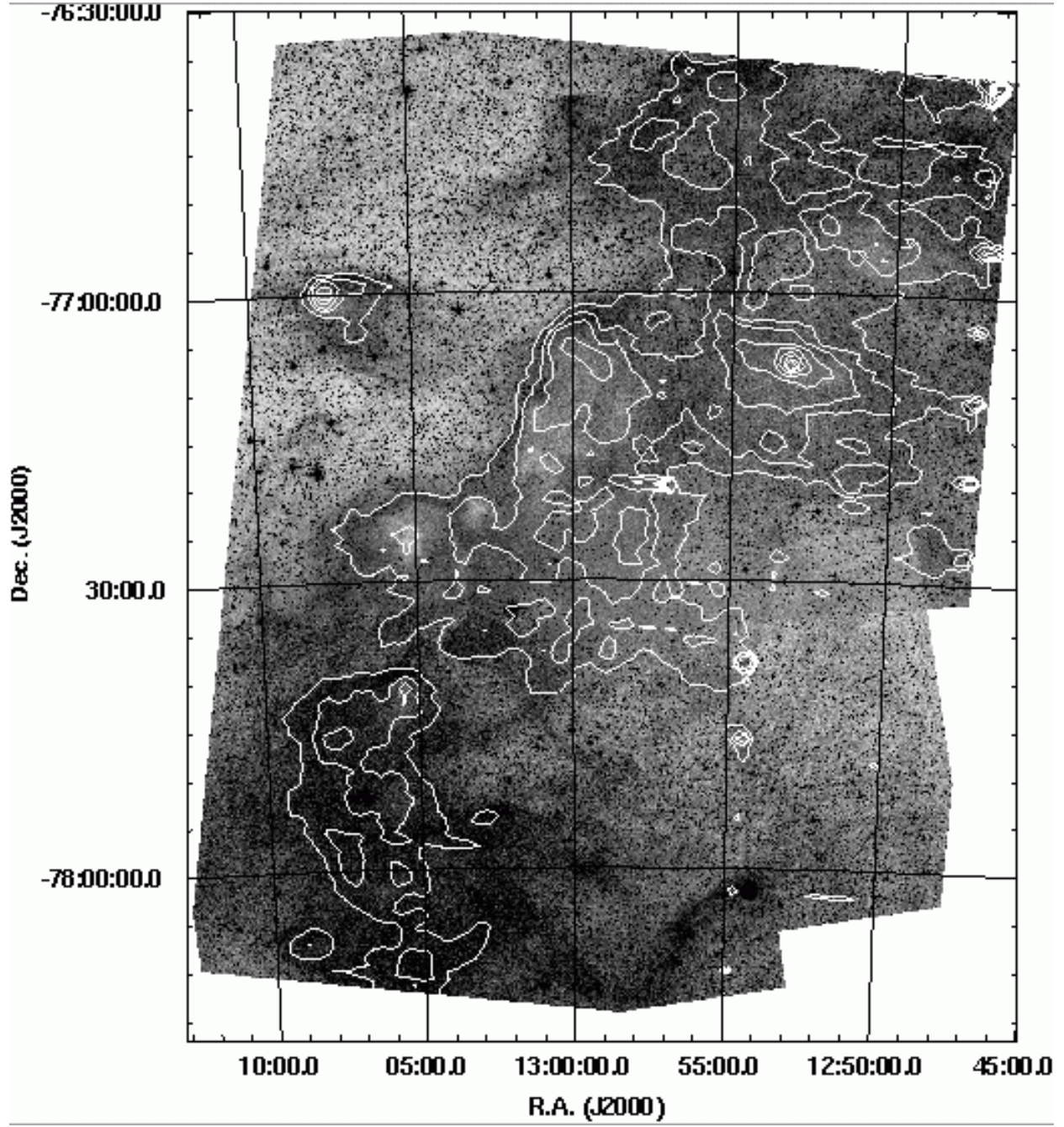} \figcaption{\label{160} Grayscale DSS R image of
Cha II overlayed with MIPS 160 \micron\ contours. Contour levels are
2, 2.5, and 3 to 8 mJy arcsec$^{-2}$ increasing by 1 mJy
arcsec$^{-2}$. The small round contours on the right edge are caused
by noisy ends of MIPS scan legs. The peaks are, from East to West,
IRAS 13036/BHR86 ($\alpha$ = 13\h\ 07\m, $\delta$ = $-$77\degree),
IRAS 12553 ($\alpha$ = 13\h, $\delta$ = $-$77\degree\ 07\am), and IRAS
12496 ($\alpha$ = 12\h\ 53\m, $\delta$ = $-$77\degree\ 07\am).}
\end{figure}

	Figure \ref{160} shows the MIPS 160 \micron\ contours of Cha
II overlaid on an Digital Sky Survey (DSS) R image of the region. The
overall shape of the Cha II cloud is outlined well at both
wavelengths. The MIPS 160 \micron\ contours reveal details about the
shape of the dense cores surrounding the bright IRAS sources, IRAS
12496-7650 (DK Cha, an Ae star) and IRAS 12553, at the center of the
cloud. The small peak to the east of the main Cha II cloud is BHR 86,
a dense core known to be harboring protostars (IRAS 13036-7644;
\citealt{leh03}). The orientation of the MIPS observations allowed BHR
86 to be observed with Cha II. However, BHR 86 was also the focus of
targeted observations by the c2d team with IRAC and MIPS and will be
discussed in detail in a future paper. The 160 \micron\ contours also
show gaps in the extended emission in the northern part of the cloud
creating ring-like structures. The most pronounced ring in the
northeast, near $\alpha$ = 12\h\ 55\m, $\delta$ = $-$76\degree\ 45\am,
contains two classical T Tauri stars (SZ 46N and SZ 47;
\citealt{che97}). Therefore, outflows from these objects may have
contributed to the formation of the ring structure. Another prominent
feature in the extended emission is the gap at $\delta$ =
$-$77\degree\ 37\am\ separating Cha II into two regions.  A group of T
Tauri stars sits on the eastern edge of the cloud near the declination
of the gap (see Fig.\ \ref{av_yso}) again suggesting that material may
have been blown out of this area by forming stars.

	The MIPS 160 \micron\ map of Cha II was also compared to the
C$^{18}$O ($J=1-0$) contours of \citet{miz99}. The peaks in the 160
\micron\ emission are well matched by the peaks of C$^{18}$O,
especially near IRAS 12496 and IRAS 12553. The C$^{18}$O map, like the
160 \micron\ emission, shows a gap in the cloud near $\delta$ =
$-$77\degree\ 37\am. There is a hint of the ring-like structures
visible in the northern part of the MIPS map in the C$^{18}$O
contours, but they are not easy to distinguish.

\subsection{SIMBA Results}\label{sest_res}

\begin{figure}
\centering
\leavevmode
   \includegraphics{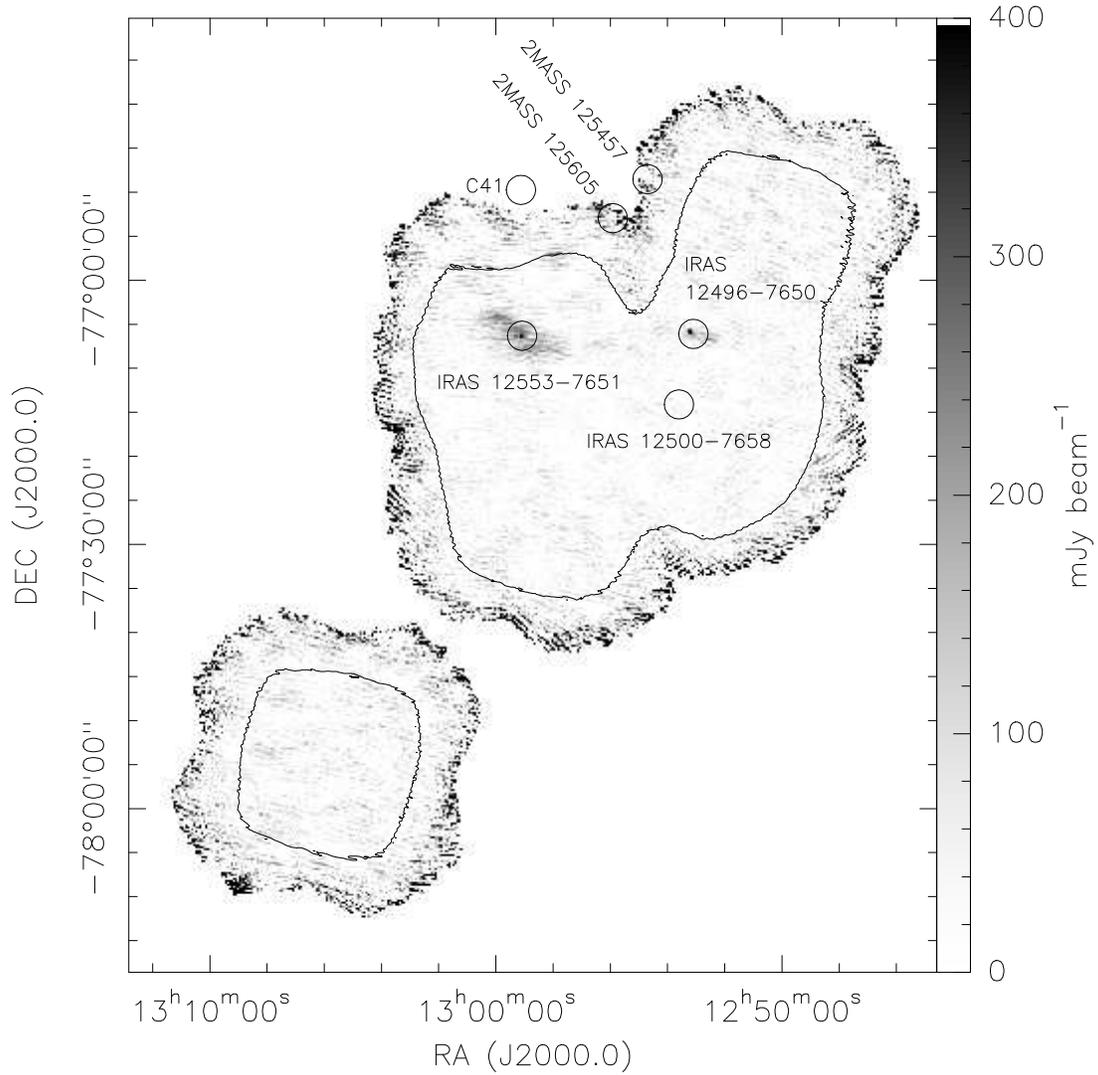}
\vskip 4.00in \figcaption{1.2 mm dust continuum emission map of the
\mbox{Cha II} complex as observed by SIMBA. The contour delimits the
area mapped with an RMS noise level below $70 ~ \rm mJy ~
beam^{-1}$. The circles are centered on the positions of the MIPS
sources discussed in detail in the text. Emission at a significant
level is only detected in the regions associated with IRAS 12496 and
IRAS 12553.\label{fig:SIMBA}}
\end{figure}

Figure \ref{fig:SIMBA} shows the dust continuum emission of the
\mbox{Cha II} complex observed by SIMBA. The RMS noise level is $32 ~
\rm mJy ~ beam^{-1}$ in the center of the map, and towards half of the
mapped area the noise level is below $70 ~ \rm mJy ~
beam^{-1}$. Emission at a significant level ($\rm SNR > 3$) is only
detected towards the regions associated with IRAS 12496 and IRAS
12553.

We identify the unresolved source at the highest intensity peak in the
millimeter dust emission maps ($694 \pm 42 ~ \rm mJy ~ beam^{-1}$)
with IRAS 12496; the MIPS data show no other bright nearby
sources. The diffuse, second highest peak in the millimeter continuum
maps ($391 \pm 48 ~ \rm mJy ~ beam^{-1}$) is near IRAS 12553, but
offset from its MIPS position by 18\arcsec. This offset appears to be
marginally significant when compared to the pointing errors toward
IRAS 12496. However, we cannot exclude that the scans towards IRAS
12553 suffer from additional pointing errors. We therefore identify
the mm peak with IRAS 12553, as there are no other bright nearby MIPS
sources. Table \ref{sourcetab} lists the 1 mm fluxes for several MIPS
point sources that we discuss in Section \ref{int} below as well as
IRAS 12496. The fluxes were derived by integration over an aperture of
80\arcsec\ diameter; the listed uncertainties reflect the statistical
noise. Possible pointing errors will not significantly affect these
fluxes, as the aperture is significantly larger than the expected
errors. The integrated flux derived for IRAS 12553 is only marginally
affected by the uncertain source position and drops by 220 mJy when
the aperture is centered on the mm peak. Surprisingly, IRAS 12553 was
not detected in the mm observations by Henning et al.\ (1993). Even
when taking the extended emission surrounding IRAS 12553 into account,
their chopped observations should have detected a source. No obvious
explanation for the differences in the inferred fluxes exists.

Extended emission is found west of IRAS 12496 with a total flux of
$3.4 \pm 0.3 ~ \rm Jy$, and surrounding IRAS 12553 with $35 \pm 1 ~
\rm Jy$. Assuming a dust temperature of $10 ~ \rm K$ and a dust
opacity of $0.01 ~ \rm cm^2 ~ g^{-1}$ at a wavelength of $1.2 ~ \rm
mm$ (\citealt{oss94}; we assume a gas-to-dust ratio of 100), we derive
a gas mass of $6.5 \pm 0.6 \, M_{\odot}$ for the extended gas
associated with IRAS 12496, and a mass of $67 \pm 2 \, M_{\odot}$ for
the gas associated with IRAS 12553. We identify these extended
features with the $\rm C^{18}O$ cores 14 (for the region close to IRAS
12496) and 19 (for IRAS 12553) found by \citet{miz99}, for which they
quote masses of $36 \, M_{\odot}$ and $64 \, M_{\odot}$,
respectively. While for the extended gas towards IRAS 12553 the mass
derived from the continuum emission and the mass derived from $\rm
C^{18}O$ agree, there is a significant difference between the mass
estimates for the extended gas near IRAS 12496. The different mass
recovery fractions could be due to the effect of spatial filtering. If
the gas near IRAS 12553 is in more compact structures than the gas
near IRAS 12496, then a larger fraction of its mass will be recovered;
gas significantly more extended than that near IRAS 12553 would be
filtered out by the data reduction and would not show up in our
maps. The different mass recovery fractions could in principle also be
due to differences in the dust temperatures. If one assumes that the
dust near IRAS 12496 is cooler than the dust near IRAS 12553, then the
difference in their mass recovery fractions is smaller than when
assuming a common dust temperature. To give an example, a common mass
recovery fraction of 50\% is inferred when assuming temperatures of $6
\rm ~ K$ for dust near IRAS 12496 and $16 \rm ~ K$ for dust near IRAS
12553. Our Spitzer images do not allow us to confirm such temperature
differences, as the calibration for extended emission is currently
uncertain.

The lack of additional detections of compact emission suggests that
only a small fraction of the gas in the \mbox{Cha II} complex ($1250
\, M_{\odot}$ as estimated by \citet{bou98} from $\rm ^{12}CO$
observations) is in compact structures with column densities exceeding
10$^{22}$ cm$^{-2}$ (corresponding to 60 mJy beam$^{-1}$, the typical
noise level in our maps). This is in line with Mizuno et al. (1999),
who speculate that the low column densities in Cha II are responsible
for the observed low star-formation activity.

\section{Selected Sources in Cha II}\label{int}

	In this section, we highlight a few interesting sources,
mainly YSOs, in the Cha II cloud. Most of the sources were originally
selected from the MIPS data because they are brighter at 70 \micron\
than 24 \micron. 2MASS 12560549-7654106 and 2MASS 13125238-7739182
were selected from Figure \ref{k24} as previously unknown YSO
candidates as described in Section \ref{colors}. C41 was chosen because
it was recently noted in the literature by \citet{bar04}.

We present full SEDs, including c2d data from IRAC (Porras et al.\
2005, in preparation), and some basic physical properties of each
source.  MIPS positions and fluxes for these sources are listed in
Table \ref{yso} and the positions are also marked as black squares on
Figure \ref{av_yso}.  SEDs including optical, 2MASS \citep{cut03},
IRAC, ISO \citep{per03}, IRAS \citep{iras}, MIPS, and millimeter
continuum (as described in this work and \citealt{hen93}) data where
available are shown in Figure \ref{sed}, and the fluxes are listed in
Table \ref{sourcetab}. We calculate bolometric luminosities for the
sources based on the SEDs shown assuming that the sources are at the
distance of Cha II.

The complementary optical data in the R, I, and z bands were obtained
using the wide-field imager at the ESO 2.2 m telescope at La Silla,
Chile. The data reduction and photometric calibration were performed
as described in Alcal\'{a} et al.\ (2004). The photometric calibration
for the z-band was performed using A0-type standards observed at an
airmass very close to 1.  The typical seeing was on the order of
0.9\arcsec. The source extraction was performed both using DAOPHOT
with PSF methods within the IRAF environment and SExtractor (Bertin \&
Arnouts 1996) with consistent results.  More details on these
observations will be published in a paper elsewhere (Alcal\'{a} et
al.\ 2005, in preparation).

\subsection{IRAS 12500-7658 / SSTc2d J125342.8-771512}

	IRAS 12500-7658 \citep{pru92} was previously detected by 2MASS
(12534285-7715114) and is also visible with IRAC and MIPS. Our SIMBA
1.2 mm map was not sensitive enough to detect IRAS 12500. However,
\citet{hen93} detected the source at 1.3 mm and classified it as an
embedded object. \citet{che97} found this source has a bolometric
temperature, $T_{bol}$, of 94 K, classifying it as a Class I object
according to the scheme of \citet{che95} where Class I objects have 70
K $< T_{bol} <$ 650 K. This source is an outlier on the 2MASS-MIPS
color plots (Fig.\ \ref{k24} \& \ref{hk24}), because it has a very red
(K$_s-$[24]) color of 9.6 mag. The only source with a similarly large
(K$_s-$[24]) color is a Class I object (IRAS 12553; see Section
\ref{iso-28}). IRAS 12500 also has a redder ([24]$-$[70]) color (3.3
mag) than most of the 24 \micron\ bright sources so that it is closer
to Class I objects (IRAS 12553 and IRAS 13036) in the [24] vs.\
([24]$-$[70]) color-magnitude diagram (Fig.\ \ref{24-70}) than the
bluer Class II objects. Further, IRAS 12500 has a positive
$\alpha_{24/K}$ (0.79) also indicating it is a Class I
object. Therefore, the MIPS data are consistent with previous
classifications of IRAS 12500 as a Class I protostar. We calculate a
bolometric luminosity, $L_{bol}$, from the SED in Figure \ref{sed} of
0.5 \lsun\ for IRAS 12500.

\subsection{2MASS 12545753-7649400 / SSTc2d J125457.5-764940}

	2MASS 12545753-7649400 does not have an IRAS source associated
with it, but was detected in all IRAC bands as well as MIPS bands 1
and 2. It has a rising SED as illustrated in Figure
\ref{sed}. However, 2MASS 125457 has a K$_s$ magnitude of 14.2,
placing it within the galaxy parameter space on the K$_s$ versus
(K$_s-$[24]) plot (Fig.\ \ref{k24}). This source also has a [24]
magnitude of 6.8, which is fainter than most YSOs but consistent with
galaxies (Fig.\ \ref{24-70}). Further, it appears slightly elongated
in the IRAC images and clearly looks like a galaxy in the I band
image. We conclude that this source is a galaxy, but include it here
to illustrate the difficulty of distinguishing YSOs from galaxies
based solely on rising SEDs in the mid-infrared. Other diagnostics
with data from different wavelengths are needed, such as the
color-magnitude diagrams presented here, to begin to separate galaxies
from YSOs in MIPS data.

\subsection{IRAS 12553-7651 / SSTc2d J125906.6-770740}\label{iso-28} 

	IRAS 12553 (ISO-ChaII-28) is one of the brightest FIR sources
in Cha II and was the target of a near-IR and ISOCAM study by
\citet{per03}. It was also detected by 2MASS (12590656-7707401), IRAS
(12553-7651; \citealt{gau92}), and at 1.2 mm with SIMBA (see Section
\ref{sest_res}). \citet{che97} reported this source has $T_{bol}$ = 99
K, classifying it as a Class I object. \citet{per03} also classified
this source as a Class I protostar based on its spectral index
($\alpha$ = 1) with $L_{bol}$ = 1.49 \lsun. We find $L_{bol}$ = 1.8
\lsun\ for IRAS 12553 from the SED in Figure \ref{sed}. The MIPS data
confirm the previous classifications of IRAS 12553. This source, like
IRAS 12500, has a redder ([24]$-$[70]) color (3.5 mag) than known
Class II sources (see Fig.\ \ref{24-70}). IRAS 12553 also has a very
red (K$_s-$[24]) color (9.5 mag) in Figure \ref{k24}. Further, we also
find a positive spectral index for this source ($\alpha_{24/K}$ =
0.76) classifying it as a Class I source.

\subsection{C41 / SSTc2d J125909.7-765104}
	
\begin{figure}
\plotone{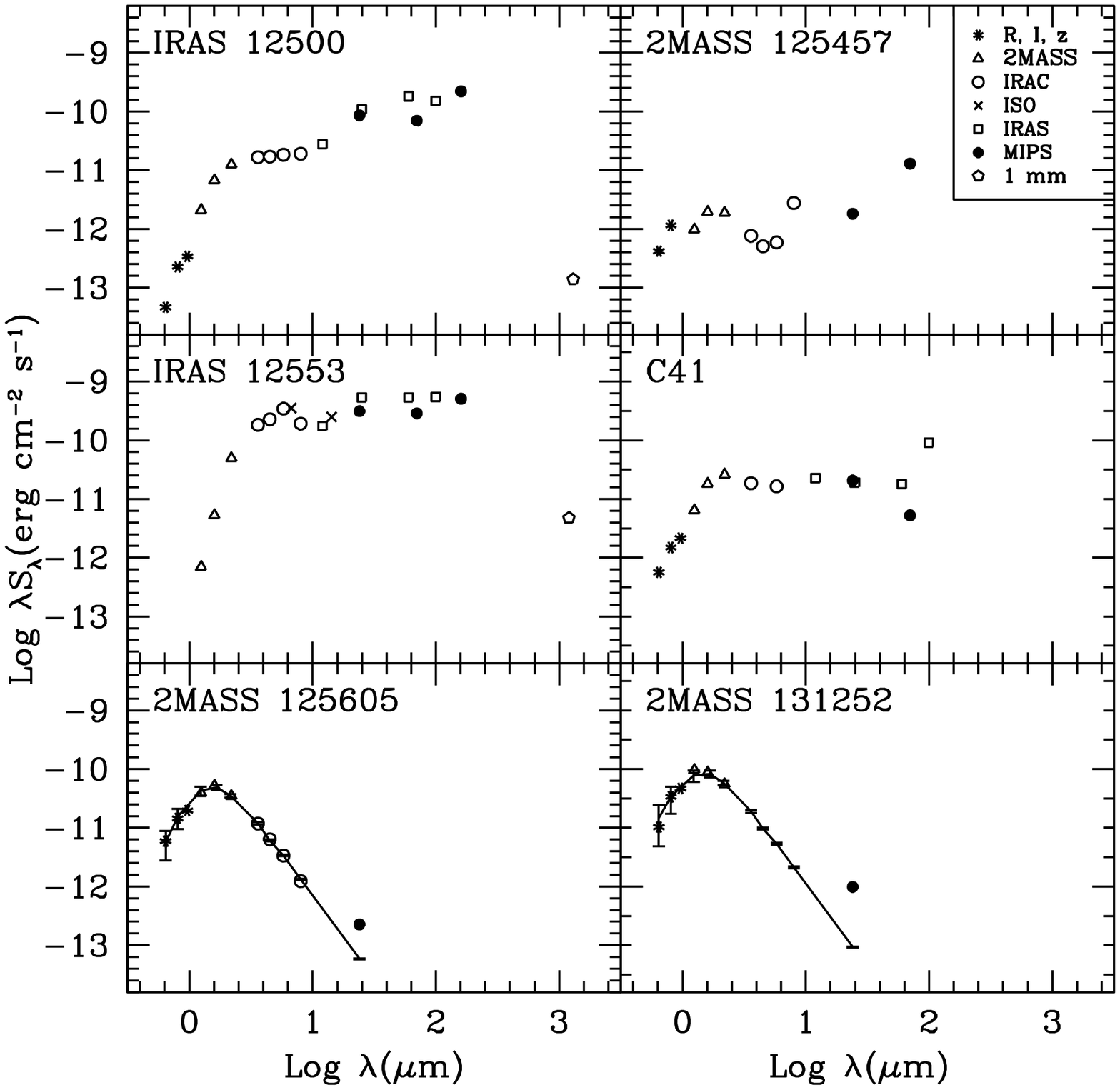} \figcaption{\label{sed} SEDs of interesting
sources discussed in Section \ref{int} including optical data, 2MASS,
IRAC, ISO \citep{per03}, MIPS, IRAS (\citealp{pru92, gau92}), 1 mm
(IRAS 12500: \citealt{hen93}, IRAS 12553: this work). A key to the
symbols is given in the figure. In all cases, the error bars are equal
to or smaller than the size of the symbol. The IRAS points in the C41
SED are for IRAS 12554-7635 which may or may not be associated with
C41. The solid line and error bars in the 2MASS 125605 and 2MASS
131252 panels are stellar models. The data show excesses over the
models at 24 \micron\ potentially indicating the presence of disks.}
\end{figure}

	C41 was recently proposed to be a classical T Tauri object,
with a spectral type of M5.5, near the sub-stellar boundary by
\citet{bar04}. They observed broad H$\alpha$ emission, a variety of
forbidden lines, and Li I absorption toward C41, which was originally
identified as a low-mass YSO candidate by \citet{vuo01}.  It has been
detected by 2MASS (12590984-7651037), IRAC, and MIPS. C41 is also
possibly associated with IRAS 12554-7635 (plotted with the SED for C41
in Figure \ref{sed}), which has a $T_{bol}$ of 81 K \citep{che97} and
was classified by \citet{gau92} as a Class I source.  However, the
IRAS 60 and 100 \micron\ fluxes for IRAS 12554 are significantly
larger than the MIPS 70 \micron\ flux of C41 (Fig.\ \ref{sed} and
Table \ref{sourcetab}) and may be confused.  Including the IRAS fluxes
in the SED, $L_{bol}$ = 0.10 \lsun, and without the IRAS fluxes C41
has an $L_{bol}$ of 0.08 \lsun. C41 has a very red (K$_s-$[24]) color
(7.3 mag) but its color lies between known Class I and Class II
sources in Figure \ref{k24}.  In Figure \ref{24-70}, C41 is in the
same region as known Class II sources, because its ([24]$-$[70]) color
is only 2.1 mag.

\subsection{2MASS 12560549-7654106 / SSTc2d J125605.5-765411}

	2MASS 12560549-7654106 was identified by its K$_s$ magnitude
(11 mag) and (K$_s-$[24]) color (2.0 mag) in Figure \ref{k24} as a
potential YSO candidate. It also has an $\alpha_{24/K} = -2.1$ which
suggests it is a Class III source. However, the K$_s$ magnitude of 11
is fainter than that of typical T Tauri stars at the distance of Cha
(K$_s$ = 8.5 -- 9.5 mag; \citealt{gau92}). 2MASS 125605 could not be
found in a search of the literature for known YSOs. It was observed by
IRAC, 2MASS, and in the optical, and has an $L_{bol}$ of only 0.05
\lsun. We compared the SED to a stellar model SED for R to 24 \micron.
The model is of a star extincted by about A$_V$ = 6 mag and uses the
dust of \citet{wei01} with a visual extinction to reddening ratio of
$R_V$ = 5.5. The SEDs of 2MASS 125605 and the model are plotted in
Figure \ref{sed}. The error bars on the model include uncertainties in
the A$_V$ and in the spectral type of the star. The SED of 2MASS
125605 is well matched by the stellar SED between R and 8 \micron.
However, the 24 \micron\ flux exceeds the stellar model prediction by
almost a factor of 4. This possible excess and the star's projected
position on the cloud suggest that 2MASS 125605 may be a previously
unknown YSO with a disk in Cha II. Further observations are needed 
to confirm its age and excess.

\subsection{2MASS 13125238-773918 / SSTc2d J131252.3-773918}

	2MASS 13125238-7739182 was also identified from the
2MASS--MIPS color magnitude plot (Fig. \ref{k24}) as a potential YSO
with (K$_s-$[24]) = 3.1 mag and K$_s$ = 10.5 mag, but has no previous
identification in the literature. This source is also fainter at K$_s$
than typical T Tauri stars (K$_s$ = 8.5 -- 9.5 mag; \citealt{gau92}).
Its spectral index, $\alpha_{24/K} = -1.68$, classifies 2MASS 131252
as a Class II object. This source was outside the c2d IRAC observation
area in Cha II, so only optical, 2MASS, and MIPS data are
available. From its SED, we calculate a low luminosity for 2MASS
131252 of $L_{bol}$ = 0.1 \lsun. We also compared the SED of this
source with that of a model of an extincted star. The 24 \micron\ flux
of 2MASS 131252 is more than a factor of 10 greater than that expected
from a star with A$_V$ of about 6 mag.  This excess suggests that
2MASS 131252 may be a previously unknown YSO with a disk. As with
2MASS 125605, further observations are needed to confirm this
conclusion.

\section{Summary}

We have presented a map of the Cha II molecular cloud at 24, 70, and
160 \micron\ as observed by the Spitzer Space Telescope instrument
MIPS (Fig.\ \ref{map}) and described the c2d Legacy team's data
processing in detail including removal of artifacts, mosaicking,
source extraction, and bandmerging.  We detected over 1500 sources at
24 \micron\ in an area of a little over 1.5 square
degrees. 

Bandmerging the source list with 2MASS and the resulting
color-magnitude diagram (Fig.\ \ref{k24}) allowed for identification
of some of the sources. Galaxies are found to be defined mostly by
faint K$_s$ magnitudes ($>$ 13 mag). Forty-four sources were found to
have K$_s$ $<$ 13 mag and (K$_s-$[24]) $>$ 1 mag and identified as
potential YSOs.  A (H$-$K$_s$) versus (K$_s-$[24]) color-color diagram
(Fig.\ \ref{hk24}) was found to be ineffective in identifying YSOs
because of confusion with galaxies. In a [24] versus ([24]$-$[70])
color-magnitude plot (Fig.\ \ref{24-70}), the data were compared with
models of YSOs from \citet{whi03} and \citet{you05} and found to be
bluer than predicted by the models. Calculating a spectral index after
the method of Lada (1987), revealed that the same sources identified
as YSO candidates in the K$_s$ vs.\ (K$_s-$[24]) diagram have
$\alpha_{24/K} > -$2.5, providing more evidence that they are YSOs and
confirming their identification in the literature. Of these 44 YSO
candidates, 4 have not been previously identified in the literature as
YSOs or candidates. The small number of new YSO candidates found with
the improved sensitivity of MIPS over previous surveys suggests that
most of the YSOs in Cha II are already known. However, future work 
combining IRAC and MIPS data may reveal more new low luminosity
YSOs.

A SIMBA dust continuum emission map of Cha II revealed significant
emission towards the regions of IRAS 12496 and IRAS 12553 only. The
lack of further detections suggests that only a small fraction of the
gas is in compact structures with high column density. The extended
emission at 160 \micron\ was found to be similar to IRAS and C$^{18}$O
\citep{miz99} maps of Cha II.

The SEDs of several YSOs associated with Cha II are presented
including the Class I sources IRAS 12553 and IRAS 12500 (Fig.\
\ref{sed}). The SEDs of two new potential YSOs, 2MASS 125605 and 2MASS
131252, were compared with a stellar model and found to have excesses
at 24 \micron\ indicating the likely presence of a disk. Follow-up
observations are needed to confirm the classification of these
previously unknown sources.

Future work will combine the MIPS and SIMBA data sets with optical,
near- and mid-infrared, including IRAC, observations of Cha II in
order to study star formation in the cloud in greater depth. The 
combined data sets will also allow for better identification and
classification of the protostellar and young stellar populations in
Cha II.

\section{Acknowledgments}

We would like to thank M. Vuong for providing her extinction map of
Cha II. Support for this work, part of the Spitzer Legacy Science
Program, was provided by NASA through contracts 1224608, 1230782, and
1230779 issued by the Jet Propulsion Laboratory, California Institute
of Technology, under NASA contract 1407. KEY was supported by NASA
under Grant No. NGT5-50401 issued through the Office of Space
Science. Astrochemistry in Leiden is supported by a NWO Spinoza grant
and a NOVA grant. This publication makes use of data products from the
Two Micron All Sky Survey, which is a joint project of the University
of Massachusetts and the Infrared Processing and Analysis
Center/California Institute of Technology, funded by NASA and the
National Science Foundation.

%%%%%%%%%%%%%% References %%%%%%%%%%%%%%%%%%%%%%

%%%%%%%%%%%%%%%%%%% Figures %%%%%%%%%%%%%%%%%%

%%%%%%%%%%%%%%%%%% Tables %%%%%%%%%%%%%%%%%%%%%

\clearpage
\begin{deluxetable}{lc}
\tablecolumns{2}
\tablecaption{Zero Point Fluxes\label{zero}}
\tablewidth{0pt}
\tablehead{
\colhead{$\lambda$} &
\colhead{$S_\nu$(0)} \\
\colhead{(\micron)} &
\colhead{(Jy)}
}

\startdata

24 & 7.24 \\
70 & 0.81  

\enddata

\end{deluxetable}

\clearpage
\begin{deluxetable}{lccccccc}
\tabletypesize{\footnotesize}
\tablecolumns{8}
\tablecaption{YSOs and YSO Candidates in Cha II\label{yso}}
\rotate
\tablewidth{0pt} 
\tablehead{
\colhead{c2d Name}                &
\colhead{Other Name(s)} 		&
\colhead{RA (J2000)\tablenotemark{a}}  &
\colhead{Dec (J2000)\tablenotemark{a}}     &
\colhead{24\micron($\sigma$)\tablenotemark{b}}  &    
\colhead{70\micron($\sigma$)\tablenotemark{b}}  &
\colhead{$\alpha_{24/K}$}  &
\colhead{Ref.\tablenotemark{c}}  \\
\colhead{}                &
\colhead{} 		&
\colhead{(\h\ \m\ \s)}  &
\colhead{(\degree\ \arcmin\ \arcsec)}     &
\colhead{(mJy)}  &
\colhead{(mJy)}  & 
\colhead{}  &  
\colhead{}  
}

\startdata 

SSTc2d J124505.8-772014 & IRAS12416-7703	 	& 12 45	07.0	& $-$77 20 13.8	& 1590(100) 	& 102(8.3)  & $-$1.29 & 1  \\
SSTc2d J124825.7-770637 & IRAS12448-7650	 	& 12 48	25.6	& $-$77 06 36.9	& 331(2.8)  	&  --       & $-$2.18 & 2  \\
SSTc2d J125217.2-780038 & IRASF12486-7744 		& 12 52 17.5    & $-$78	00 37.4 & 29.6(0.23)	&  --	    & $-$2.44 & 2  \\ 
SSTc2d J125230.6-771513 & IRASF12488-7658/C13	 	& 12 52	30.5	& $-$77 15 13.3	& 34.7(0.20)	&  --       & $-$2.33 & 3  \\
SSTc2d J125342.8-771512 & IRAS12500-7658	 	& 12 53	42.6	& $-$77 15 10.3	& 687(4.2)  	& 1630(100) &    0.79  & 2  \\
SSTc2d J125605.5-765411 & 2MASS12560549-7654106 	& 12 56	05.2	& $-$76 54 10.6	& 1.81(0.09)	&  --       & $-$2.10  & 4  \\
SSTc2d J125633.6-764545 & Sz46N		 		& 12 56	33.6	& $-$76 45 45.6	& 57.1(0.28)	&  --       & $-$1.16 & 5  \\
SSTc2d J125658.7-764707 & Sz47				& 12 56	58.5	& $-$76 47 07.2	& 1.11(0.07)	&  --       & $-$1.84 & 5  \\
SSTc2d J125711.7-764011 & IRAS12535-7623/CHIIXR2 	& 12 57	11.5	& $-$76 40 10.5	& 430(3.5)  	& 147(12)   & $-$0.84 & 2  \\
SSTc2d J125806.8-770909 & ISO-ChaII-13		 	& 12 58	06.7	& $-$77 09 09.6	& 8.39(0.14)	&  --       & $-$0.87 & 6  \\
SSTc2d J125906.6-770740 & IRAS12553-7651/ISO-ChaII-28	& 12 59	06.4	& $-$77 07 40.0	& 2500(100) 	& 6720(290) &    0.76  & 6  \\
SSTc2d J125909.7-765104 & C41				& 12 59	10.4	& $-$76 51 03.1	& 165(1.2)  	& 123(9.7)  & $-$0.10  & 3  \\
SSTc2d J125926.4-774709 & IRAS12556-7731	 	& 12 59	26.3	& $-$77 47 08.8	& 294(1.2)  	&  --       & $-$1.70  & 2  \\
SSTc2d J130053.1-765416 & ISO-ChaII-55/Sz49/CHIIXR9	& 13 00	53.2	& $-$76 54 14.9	& 132(0.86) 	& 131(8.8)  & $-$0.47 & 6  \\
SSTc2d J130053.3-770909 & ISO-ChaII-51/Sz48/CHIIXR7	& 13 00	53.4	& $-$77 09 08.9	& 96.6(0.40)	&  --       & $-$1.06 & 6  \\
SSTc2d J130055.3-771022 & ISO-ChaII-52/Sz50/CHIIXR8	& 13 00	55.5	& $-$77 10 20.3	& 416(2.2)  	& 604(31)   & $-$0.68 & 6  \\
SSTc2d J130059.3-771403 & ISO-ChaII-54/CHIIXR10/C48 	& 13 00	59.5	& $-$77 14 03.1	& 566(3.4)  	& 288(15)   & $-$0.87 & 6, 3  \\
SSTc2d J130158.9-775122 & Sz51				& 13 01	58.4	& $-$77 51 19.9	& 123(0.57) 	& 109(8.4)  & $-$0.99 & 5  \\
SSTc2d J130213.5-763758 & IRAS12584-7621/CM Cha		& 13 02	13.9	& $-$76 37 57.8	& 416(2.1)  	& 411(22)   & $-$0.81 & 2  \\
SSTc2d J130222.8-773449 & C50				& 13 02	22.8	& $-$77 34 49.7	& 8.07(0.10)	&  --       & $-$1.34 & 3  \\
SSTc2d J130247.5-770240 & IRAS12589-7646	 	& 13 02	47.6	& $-$77 02 40.1	& 209(2.0)  	&  --       & $-$2.16 & 2  \\
SSTc2d J130423.9-765002 & Hn23				& 13 04	23.6	& $-$76 50 03.3	& 226(4.1)  	3& 448(24)   & $-$0.96 & 5  \\
SSTc2d J130424.9-775230 & Sz52				& 13 04	24.9	& $-$77 52 30.6	& 49.0(0.19)	&  --       & $-$0.89 & 5  \\
SSTc2d J130455.7-773950 & Hn24				& 13 04	55.9	& $-$77 39 49.5	& 280(1.9)  	& 243(13)   & $-$0.82 & 5  \\
SSTc2d J130508.5-773343 & Hn25/C53			& 13 05	08.4	& $-$77 33 42.9	& 79.0(0.90)	&  --       & $-$0.96 & 5, 3  \\
SSTc2d J130512.7-773052 & Sz53			 	& 13 05	12.5	& $-$77 30 52.9	& 91.6(0.66)	&  --       & $-$0.90  & 5  \\
SSTc2d J130520.6-773902 & Sz54				& 13 05	20.8	& $-$77 39 00.9	& 299(1.4)  	& 284(14)   & $-$1.30  & 5  \\
SSTc2d J130630.3-773401 & Sz55				& 13 06	30.4	& $-$77 34 00.4	& 29.4(0.21)	&  --       & $-$0.99 & 5  \\
SSTc2d J130638.5-773036 & Sz56				& 13 06	38.7	& $-$77 30 35.6	& 56.7(0.22)	&  --       & $-$0.91 & 5  \\
SSTc2d J130656.5-772310 & Sz57/C60			& 13 06	56.5	& $-$77 23 09.7	& 36.7(0.17)	&  --       & $-$1.33 & 5, 3  \\
SSTc2d J130657.4-772342 & IRAS13030-7707/C61/Sz58	& 13 06	57.2	& $-$77 23 39.6	& 386(1.9)  	& 636(39)   & $-$0.75 & 3  \\
SSTc2d J130709.0-773031 & Sz59				& 13 07	09.1	& $-$77 30 30.7	& 265(1.5)  	&  --       & $-$1.05 & 5  \\
SSTc2d J130718.0-774053 & C62				& 13 07	18.0	& $-$77 40 53.2	& 14.3(0.10)	&  --       & $-$1.02 & 3  \\
SSTc2d J130722.8-773724 & Sz60				& 13 07	22.4	& $-$77	37 22.6 & 63.8(0.74)    &  --       & $-$1.19 & 5  \\
SSTc2d J130748.5-774121 & Hn26				& 13 07	48.4	& $-$77 41 21.7	& 68.1(0.27)	&  --       & $-$0.98 & 5  \\
SSTc2d J130806.2-775505 & Sz61				& 13 08	06.4	& $-$77 55 04.9	& 724(4.2)  	& 572(27)   & $-$0.79 & 5  \\
SSTc2d J130827.2-774323 & C66				& 13 08	27.2	& $-$77 43 23.5	& 5.85(0.09)	&  --       & $-$1.13 & 3  \\
SSTc2d J130910.3-770944 & IRASF13052-7653/CHIIXR60      & 13 09	09.9	& $-$77	09 43.8 & 122(1.6)      &  --       & $-$0.89 & 2  \\
SSTc2d J130950.3-775724 & Sz62				& 13 09	50.3	& $-$77 57 24.3	& 130(0.88) 	&  --	    & $-$1.06 & 5  \\
SSTc2d J131004.3-771045 & Sz63				& 13 10	04.2	& $-$77 10 44.9	& 42.1(1.1) 	&  --	    & $-$1.04 & 5  \\
SSTc2d J131025.3-772909 & 2MASS13102531-7729085         & 13 10	25.3	& $-$77	29 08.6 & 3.84(0.10)    &  --       & $-$1.13 & 4  \\
SSTc2d J131103.3-765333 & 2MASS13110329-7653330		& 13 11 03.3    & $-$76 53 33.1 & 1.56(0.11)    &  --       & $-$1.71 & 4  \\
SSTc2d J131252.3-773918 & 2MASS13125238-7739182		& 13 12	52.3	& $-$77 39 18.1	& 7.89(0.08)	&  --       & $-$1.68 & 4  \\
SSTc2d J131403.5-775308 & Sz64				& 13 14	03.6	& $-$77 53 07.9	& 37.2(0.18)	&  --	    & $-$0.91 & 5   
														        
\enddata													        
														        
\tablenotetext{a}{Positions are from the 24 \micron\ map.}							        
\tablenotetext{b}{$\sigma$ is the flux uncertainty calculated using c2dphot as described in 			        
Section \ref{extract}. The absolute uncertainty in the flux is closer to 10\% at 24 \micron\ 			        
and 20\% at 70 \micron.} 
\tablenotetext{c}{1: \citet{gau92}, 2: \citet{pru92}, 3: \citet{vuo01}, 4: This work, 5: \citet{che97}, 6: \citet{per03}}

\end{deluxetable}

\clearpage
\begin{deluxetable}{lccc}
\tabletypesize{\footnotesize}
\tablecolumns{4}
\tablecaption{SEDs of Selected Sources in Cha II\label{sourcetab}}
%\rotate
\tablewidth{0pt} 
\tablehead{
\colhead{Name(s)}                &
\colhead{$\lambda$} &
\colhead{Flux($\sigma$)\tablenotemark{a}} &
\colhead{Ref.\tablenotemark{b}}    \\
\colhead{}  &
\colhead{(\micron)} &
\colhead{(mJy)}  &
\colhead{}      
}

\startdata 

IRAS 12496-7650/DK Cha\tablenotemark{c} &  70 &  36100(1200) & 1 \\
     	     &		   1200 &  1470(120)\tablenotemark{d} & 1 \\
IRAS 12500-7658     & 0.643 & 0.0099(0.0009) & 1 \\
		  	& 0.805 & 0.061(0.002)  & 1 \\
		  	& 0.965 & 0.11(0.006)  	& 1 \\
		  	& 1.25  & 0.869(0.06) 	& 2  \\
		  	& 1.6   & 3.56(0.14)  	& 2 \\
		  	& 2.2   & 9.19(0.20)  	& 2 \\
		  	& 3.6   & 20.0(0.19)  	& 1 \\
		  	& 4.5   & 25.6(0.21) 	& 1 \\
		  	& 5.8   & 35.6(0.23) 	& 1 \\
		  	& 8    & 51.0(0.29)  	& 1 \\
		  	& 12   & 110(22) 	& 3 \\
		  	& 24   & 687(4.2) 	& 1 \\
		  	& 25   & 900(180)  	& 3 \\
		  	& 60   & 3630(730)  	& 3 \\
		  	& 70   & 1630(100) 	& 1 \\
		  	& 100  & 5060(1100)  	& 3 \\
		  	& 160\tablenotemark{e}  &  11800(2400) & 1 \\
		  	& 1200\tablenotemark{d} & $<$260\tablenotemark{f}  & 1 \\
		  	& 1300   & 59.7(15)  	& 4  \\
2MASS 12545753-7649400 & 0.643 & .0896(0.005) & 1 \\
		  	& 0.805 & 0.309(0.01) & 1 \\
		  	& 0.965 & 1.67(0.15)	& 1 \\
		  	& 1.25  & 0.407(0.05)	& 2  \\
		  	& 1.6   & 1.04(0.11)	& 2 \\
		  	& 2.2   & 1.37(0.12)	& 2 \\
		  	& 3.6   & 0.92(0.07)	& 1 \\
		  	& 4.5   & 0.76(0.05)	& 1 \\
		  	& 5.8   & 1.14(0.07)	& 1 \\
		  	& 8    & 7.37(0.27)	& 1 \\
		  	& 24   & 14.5(0.18)	& 1 \\
		        & 70   & 302(16) 	& 1 \\
2MASS 12560549-7654106 &  0.643 & 1.27(0.011) & 1 \\
		  & 0.805 & 3.89(0.033) & 1 \\
		  & 0.965 & 6.34(0.11)	& 1 \\
		  & 1.25  & 16.3(0.31)	& 2  \\
		  & 1.6   & 27.7(0.51)	& 2 \\
		  & 2.2   & 25.4(0.40)	& 2 \\
		  & 3.6   & 14.2(0.12)	& 1 \\
		  & 4.5   & 9.54(0.09)	& 1 \\
		  & 5.8   & 6.50(0.06)	& 1 \\
		  & 8     & 3.31(0.04)	& 1 \\
		  & 24    & 1.81(0.09)	& 1 \\
IRAS 12553-7651/ISO-ChaII-28  &  1.25  & $<$0.29	& 2  \\
		  & 1.6   & 2.83(0.11)	& 2 \\
		  & 2.2   & 36.2(0.70)	& 2 \\
		  & 3.6   & 219(3.2)	& 1 \\
		  & 4.5   & 342(9.3)	& 1 \\
		  & 5.8   & 667(5.8)	& 1 \\
		  & 6.7   & 793(26)	& 5 \\
		  & 8     & 514(10)	& 1 \\
		  & 12    & 700(70)	& 3 \\
		  & 14.3  & 1190(25)	& 5 \\
		  & 24    & 2500(100)	& 1 \\
		  & 25    & 4430(220)	& 3 \\
		  & 60    & 10600(530)	& 3 \\
		  & 70    & 6720(280)	& 1 \\
		  & 100   & 18100(5400)	& 3 \\
		  & 160\tablenotemark{e}   & 27200(2700) & 1 \\
		  & 1200\tablenotemark{d}  & 1910(130)  & 1 \\
C41		  & 0.643 & 0.121(0.002) & 1 \\
		  & 0.805 & 0.40(0.007) & 1 \\
		  & 0.965 & 0.69(0.012)	& 1 \\
		  & 1.25  & 2.66(0.12) 	& 2  \\
		  & 1.6   & $<$9.64 	& 2 \\
		  & 2.2   & $<$19.0	& 2 \\
		  & 3.6   & 22.3(0.32)	& 1 \\
		  & 5.8   & 31.7(0.22)	& 1 \\
		  & 12   & 90(18)	& 3 \\
		  & 24   & 165(1.2)	& 1 \\
		  & 25   & 160(32)	& 3 \\
		  & 60   & 360(72)	& 3 \\
		  & 70   & 123(9.7)	& 1 \\
		  & 100  & 3030(610)	& 3 \\
2MASS 13125238-7739182	 & 0.643 & 2.21(0.019) & 1 \\
		  & 0.805 & 9.08(0.24) & 1 \\
		  & 0.965 & 15.1(0.54)	& 1 \\
		  & 1.25  & 40.3(0.85)	& 2  \\
		  & 1.6   & 46.6(1.1)	& 2 \\
		  & 2.2   & 40.4(0.78)	& 2 \\
		  & 24   & 7.89(0.08)	& 1 \\

\enddata

\tablenotetext{a}{For the IRAC and MIPS 24 and 70\micron\ data,
$\sigma$ is the flux uncertainty calculated using c2dphot as described
in Section \ref{extract}. The absolute uncertainty in the flux is
closer to 10\% at 24 \micron\ and 20\% at 70 \micron. The 160 \micron\
$\sigma$ reflects a flux uncertainty of 20\%. The 1.2 mm $\sigma$
reflects the statistical noise.}

\tablenotetext{b}{1: This work, 2: \citet{cut03}, 3: \citet{iras}, 4:
\citet{hen93}, 5: \citet{per03}}

\tablenotetext{c}{IRAS 12496 is saturated at 24 and 160 \micron.}

\tablenotetext{d}{1.2 mm fluxes are aperture fluxes with a 80\arcsec\
diameter aperture.}

\tablenotetext{e}{160 \micron\ fluxes are aperture fluxes with a
40\arcsec\ diameter aperture.}

\tablenotetext{f}{2-$\sigma$ upper limit.}

\end{deluxetable}


\begin{thebibliography}{}
\bibitem[Alcal\'{a} et al.(2000)]{alc00}Alcal\'{a}, J. M., Covino, E., Sterzik, M. F., Schmitt, J. H. M. M., Krautter, J., \& Neuh\"{a}user, R. 2000, A\&A, 355, 629
\bibitem[Alcal\'{a} et al.(2004)]{alc04}Alcal\'{a}, J. M., Wachter, S.,
Covino, E., Sterzik M. F., Durisen R. H., Freyberg M., Hoard D. W., \&
Cooksey K. 2004, A\&A, 416, 677
\bibitem[Barrado y Navascu\'{e}s \& Jayawardhana(2004)]{bar04}Barrado y Navascu\'{e}s, D. \& Jayawardhana, R. 2004, ApJ, 615, 280
\bibitem[Bertin \& Arnouts(1996)]{ber96}Bertin E. \& Arnouts S. 1996, A\&AS, 117, 393
\bibitem[Boulanger et al.(1998)]{bou98}Boulanger, F., Brofman, L., Dame, T. M., \& Thaddeus, P. 1998, A\&A, 332, 273
\bibitem[Bourke et al.(1995)]{bou95}Bourke, T. L., Hyland, A. R., Robinson, G., James, S. D., \& Wright, C. M. 1995, MNRAS, 276, 1067
\bibitem[Cambr\'{e}sy(1999)]{cam99}Cambr\'{e}sy, L. 1999, A\&A, 345,965 
\bibitem[Chen et al.(1997)]{che97}Chen, H., Grenfell, T. G., Myers, P. C., Hughes, J. D. 1997, ApJ, 478, 295
\bibitem[Chen et al.(1995)]{che95}Chen, H., Myers, P. C., Ladd, E. F., \& Wood, D. O. S. 1995, ApJ, 445, 377
\bibitem[Cutri et al.(2003)]{cut03}Cutri, R. M., et al. 2003, Explanatory Supplement to the 2MASS All Sky Data Release, (Pasadena: IPAC), http://www.ipac.caltech.edu/2mass/releases/allsky/doc/explsup.html
\bibitem[Evans et al.(2003)]{eva03}Evans, N. J., II, et al. 2003, PASP, 115, 965
\bibitem[Evans et al.(2004)]{eva04}Evans, N. J., II, et al. 2004, First Delivery of Data from the c2d Legacy Project, (Pasadena: Spitzer Science Center), http://ssc.spitzer.caltech.edu/legacy/
\bibitem[Gauvin \& Strom(1992)]{gau92}Gauvin, L. S. \& Strom, K. M. 1992, ApJ, 385, 217
\bibitem[Gordon et al.(2005a)]{gor05a}Gordon, K. D., et al. 2005a, PASP, in press, astro-ph/0502079
\bibitem[Gordon et al.(2005b)]{gor05b}Gordon, K. D., et al. 2005b, astro-ph/0502080
\bibitem[Henning et al.(1993)]{hen93}Henning, Th., Pfau, W., Zinnecker, H., and Prusti, T. 1993, A\&A, 276, 129
\bibitem[Hughes \& Hartigan(1992)]{hug92}Hughes, J. \& Hartigan, P. 1992, AJ, 104, 680
\bibitem[IRAS PSC(1988)]{iras}IRAS PSC 1988, ed. C. Beichmann et al. (NASA RP-1190; Washington: GPO)
\bibitem[Knee(1992)]{kne92}Knee, L. B. G. 1992, A\&A, 259, 283
\bibitem[Lada(1987)]{lad87}Lada, C. J. 1987, in IAU Symp. 115, Star Formation Regions, ed. M. Peimbert \& J. Jugaku (Dordrecht: Reidel), 1 
\bibitem[Lehtinen \& Higdon(2003)]{leh03}Lehtinen, K. \& Higdon, J. L. 2003, A\&A, 398, 583
\bibitem[Liseau et al.(1996)]{lis96}Liseau, R. et al. 1996, A\&A, 315, L181
\bibitem[Lonsdale et al.(2003)]{lon03}Lonsdale, C. J. et al. 2003, PASP, 115, 897
\bibitem[Makovoz(2004)]{mac04}Makovoz, D. 2004, Spitzer Mosaicker Version 1.6, (Pasadena: Spitzer Science Center), http://ssc.spitzer.caltech.edu/postbcd/
\bibitem[Mardones et al.(1997)]{mar97}Mardondes, D., Myers, P. C., Tafalla, M., Wilner, D. J., Bachiller, R., \& Garay, G. 1997, ApJ, 489, 719
\bibitem[Marleau et al.(2004)]{mar04}Marleau, F. R., et al. 2004, ApJS, 154, 66 
\bibitem[Masci et al.(2005)]{mas04}Masci, F. J., Laher, R., Fang, F., Fowler, J., Lee, W., Stolovy, S., Padgett, D. \& Moshir, M. 2005, ASP Conference Series, Astronomical Data Analysis Software and Systems XIV, eds. Shopbell, P. L., Britton, M. C., \& Ebert, R. http://xxx.lanl.gov/abs/astro-ph/0411316
\bibitem[Mizuno et al.(1995)]{miz95}Mizuno, A., Onishi, T., Yonekura, Y., Nagahama, T., Ogawa, H., Fukui, Y. 1995, ApJ, 445, L161 
\bibitem[Mizuno et al.(1999)]{miz99}Mizuno, A., et al. 1999, PASJ, 51, 859
\bibitem[Neufeld, Melnick, \& Harwit(1998)]{neu98}Neufeld, D. A., Melnick, G. J., Harwit, M. 1998, ApJ, 506, L75
\bibitem[Onishi et al.(1996)]{oni96}Onishi, T., Mizuno, A., Kawamura, A., Ogawa, H., Fukui, Y. 1996, ApJ, 465, 815
\bibitem[Ossenkopf \& Henning(1994)]{oss94}Ossenkopf, V. \& Henning, Th. 1994, A\&A, 291, 943
\bibitem[Persi et al.(2003)]{per03}Persi, P., Marenzi, A. R., Gomez, M. \& Olofsson, G. 2003, A\&A, 399, 995
\bibitem[Press et al.(1997)]{pre97}Press, W. H., Teukolsky, S. A.,
Vetterling, W. T., \& Flannery, B. P. 1997, Numerical Recipes in C,
(Cambridge:Cambridge U. Press)
\bibitem[Prusti et al.(1992)]{pru92}Prusti, T, Wittet, D. C. B., Assendorp, R., \& Wesselius, P. R. 1992, A\&A, 260, 151 
\bibitem[Rieke et al.(2004)]{rie04}Rieke, G. et al. 2004, ApJS, 154, 25
\bibitem[Schechter, Mateo, \& Saha(1993)]{sch93}Schechter, P.L., Mateo, M., \& Saha, A. 1993, PASP, 105, 693 
\bibitem[Schwartz(1977)]{sch77}Schwartz, R. D. 1977, ApJS, 35, 161
\bibitem[Sivia(1996)]{siv96}Sivia, D. S. 1996, Data Analysis: A Bayesian Tutorial, (Oxford: Clarendon Press)
\bibitem[MIPS Data Handbook(2004)]{mips04}Spitzer Space Telescope Multiband Imaging Photometer for Spitzer (MIPS) Data Handbook Version 1.1, 2004, (Pasadena: Spitzer Science Center), http://ssc.spitzer.caltech.edu/mips/
\bibitem[Spitzer Observer's Manual(2004)]{spi04}Spitzer Space Telescope Observers Manual - Version 5.0, 2004, (Pasadena: Spitzer Science Center), http://ssc.spitzer.caltech.edu/documents/
\bibitem[Surace et al.(2004)]{sur04}Surace, J. A. et al. 2004, The SWIRE ELAIS N1 Image Atlases and Source Catalogs, (Pasadena: Spitzer Science Center), http://ssc.spitzer.caltech.edu/legacy/
\bibitem[Tachihara, Mizuno \& Fukui(2000)]{tac00}Tachihara, K., Mizuno, A., \& Fukui, Y. 2000, ApJ, 528, 817
\bibitem[Vuong, Cambr\'{e}sy, \& Epchtein(2001)]{vuo01}Vuong, M. H., Cambr\'{e}sy, L. \& Epchtein, N. 2001, A\&A, 379, 208
\bibitem[Wainscoat et al.(1992)]{wai92}Wainscoat, R. J., Cohen, M., Volk, K., Walker, H. J., \& Schwartz, D. E. 1992, ApJS, 83, 111
\bibitem[Whitney et al.(2003)]{whi03}Whitney, B. A., Wood, K., Bjorkman, J. E., \& Cohen, M. 2003, ApJ, 598, 1079 
\bibitem[Whittet et al.(1991)]{whi91}Whittet, D. C. B., Assendorp, R., Prusti, T., Wesselius, P. R., \& Roth, M. 1991, A\&A, 251, 524 
\bibitem[Whittet et al.(1997)]{whi97}Whittet, D. C. B., Prusti, T., Franco, G. A. P., Gerakines, P. A., Kilkenny, D., Larson, K. A., \& Wesselius, P. R. 1997, A\&A, 327, 1194 
\bibitem[Weingartner \& Draine(2001)]{wei01}Weingartner, J. C. \& Draine, B. T. 2001, ApJ, 548, 296
\bibitem[Young \& Evans(2005)]{you05}Young, C. H. \& Evans, N. J., II 2005, ApJ, in press, astro-ph/0503456
\bibitem[Yun et al.(1999)]{yun99}Yun, J. L., Moreira, M. C., Afonso, J. M., \& Clemens, D. P. 1999, ApJ, 118, 990
\end{thebibliography}
\end{document}